\documentclass[journal]{IEEEtran}

\usepackage{etoolbox}
\makeatletter
\patchcmd{\@makecaption}
  {\scshape}
  {\normalsize}
  {}
  {}
\makeatletter
\patchcmd{\@makecaption}
  {\\}
  {\normalsize{:}\ }
  {}
  {}
\makeatother
\def\tablename{\normalsize{TABLE}}

\makeatletter
\def\endthebibliography{%
  \def\@noitemerr{\@latex@warning{Empty `thebibliography' environment}}%
  \endlist
}
\makeatother
\IEEEoverridecommandlockouts

\usepackage{cite}
\usepackage{amsmath,amssymb,amsfonts}

\usepackage{algorithm}
\usepackage{array}
\usepackage[caption=false,font=footnotesize,labelfont=sf,textfont=sf]{subfig}
\usepackage{fixltx2e}
\usepackage{stfloats}
\usepackage{url}
\usepackage{algpseudocode}
\usepackage{graphicx}
\usepackage{textcomp}
\usepackage{xcolor}
\usepackage{mathrsfs}
\usepackage{float}
\usepackage{bm}
\usepackage{enumerate}
\usepackage{url}

\def\BibTeX{{\rm B\kern-.05em{\sc i\kern-.025em b}\kern-.08em
    T\kern-.1667em\lower.7ex\hbox{E}\kern-.125emX}}

\begin{document}

\title{Digital Twin Enhanced Deep Reinforcement Learning for Intelligent Omni-Surface Configurations in MU-MIMO Systems}

\author{Xiaowen~Ye, Xianghao~Yu,~\IEEEmembership{Senior~Member,~IEEE}, and
Liqun~Fu,~\IEEEmembership{Senior~Member,~IEEE}
\thanks{Xiaowen~Ye and Liqun~Fu are with the School of Informatics, Xiamen University, Xiamen, China. E-mail: \{xiaowen@stu.xmu.edu.cn, liqun@xmu.edu.cn\}. (\textit{Corresponding author: Liqun Fu.})}
\thanks{Xianghao Yu is with the Department of Electrical Engineering, City University of Hong Kong, Hong Kong, China. E-mail: \{alex.yu@cityu.edu.hk\}.}
\thanks{The work of Liqun Fu was partially supported by the National Natural Science Foundation of China (No. U23A20281), and the Open Research Project Programme of the State Key Laboratory of Internet of Things for Smart City (University of Macau) (Ref. No.: SKL-IoTSC(UM)-2021-2023/ORP/GA03/2022). The work of Xianghao Yu was supported in part by the Hong Kong Research Grants Council under Grant No. 11208724, and in part by the NSFC Young Scientists Fund No. 62301468.}
}

\renewcommand{\thefootnote}{\arabic{footnote}}
\maketitle

\begin{abstract}
Intelligent omni-surface (IOS) is a promising technique to enhance the capacity of wireless networks, by reflecting and refracting the incident signal simultaneously. Traditional IOS configuration schemes, relying on all sub-channels' channel state information and user equipments' mobility, are difficult to implement in complex realistic systems. Existing works attempt to address this issue employing deep reinforcement learning (DRL), but this method requires a lot of trial-and-error interactions with the external environment for efficient results and thus cannot satisfy the real-time decision-making. To enable model-free and real-time IOS control, this paper puts forth a new framework that integrates DRL and digital twins. As a first step, DeepIOS, a DRL based IOS configuration scheme with the goal of maximizing the sum data rate, is developed to jointly optimize the phase-shift and amplitude of IOS in multi-user multiple-input-multiple-output (MU-MIMO) systems. Thereafter, in order to further reduce the computational complexity, DeepIOS introduces an action branch architecture, which decides two optimization variables in parallel in a separate fashion. Finally, a digital twin module is constructed through supervised learning as a pre-verification platform for DeepIOS, such that the decision-making's real-time can be guaranteed. The formulated framework is a closed-loop system, in which the physical space provides data to establish and calibrate the digital space, while the digital space generates a large number of experience samples for DeepIOS training and sends the trained parameters to the IOS controller for configurations. Numerical results show that compared with random and MAB schemes, the proposed framework attains a higher data rate and is more robust to different settings. Furthermore, the action branch architecture reduces DeepIOS's computational complexity, and the digital twin module improves DeepIOS's convergence speed and run-time.
\end{abstract}


\section{Introduction}\label{intro}
\IEEEPARstart{W}{ith} the rapid development of metasurfaces, intelligent reflecting surface (IRS) has emerged as a promising technique to meet the increasing demand for sixth-generation wireless networks (6G) \cite{wu2021intelligent}. IRS is an array with a large number of reconfiguration elements, each of which introduces a reflection response for electromagnetic signals impinging on it. By properly tuning elements' phase-shift via a programmable controller, the reflected signals propagate in a desired manner, thus effectively alleviating the wireless channel fading \cite{wu2021intelligent}. The studies of IRS assisted communications are in full swing in various fields, including but not limited to the designs of maximizing sum-rate \cite{huang2023rate, wang2022intelligent, xu2023intelligent}, energy and spectral efficiency \cite{yang2021energy, niu2023active, aung2023energy, yuan2020intelligent}, and minimizing power consumption \cite{pan2020intelligent}.

Reflecting-type IRS, however, completely reflects the incident signal towards the same side at the surface, which consequently limits the services for the user equipments (UEs) on the other side. Fortunately, intelligent omni-surface (IOS) \cite{zeng2021reconfigurable}, a new instance of IRS, has emerged to achieve ubiquitous wireless communications. In contrast to IRS, IOS possesses the dual functionality of reflection and refraction \cite{zhang2022intelligent1}, in the sense that the incident signal is simultaneously reflected and refracted towards the UEs on both sides of the surface. Similar to IRS, IOS consists of many reconfiguration elements, which are able to customize the wireless propagation environment. Generally, IOS can be operated in different control protocols \cite{fang2023intelligent}, in which the energy splitting (ES) and mode selection (MS) protocols are attractive candidates owing to their easy implementation. Specifically, the ES protocol splits the signal incident upon each element into reflected and refracted signals of different energies, whereas the MS protocol divides all elements into two groups operating in reflective and refractive modes, respectively. With the diversity of available protocols, IOS provides more flexible applications compared to IRS \cite{zhang2022intelligent2}.

For a focus, this paper considers a multi-user multiple-input-multiple-output (MU-MIMO) system, where the base station (BS) serves multiple UEs simultaneously. Due to obstacles' presence, an IOS with a programmable controller helps to enhance the signals’ strength at UEs. The system may employ different protocols to manipulate the IOS; and thus apart from the phase-shift, the controller also adjusts each element's reflective and refractive amplitudes. The system’s objective is to maximize the system's sum-rate.
Several traditional optimization based approaches for IOS assisted communications have been proposed in \cite{zhang2022meta, liu2022full, zhang2021intelligent, chen2022robust, cai2022joint, wang2022intelligentomni, fang2022intelligent, benaya2023physical, wang2022safeguarding, zhang2022dual}. For example, \cite{zhang2022meta, liu2022full, zhang2021intelligent} designed various transmission schemes with sum-rate maximization by adjusting the phase-shift of IOS, while \cite{chen2022robust} and \cite{cai2022joint} jointly optimized the phase-shift and amplitude of ES-IOS to save energy required for services. To guarantee secure transmissions, \cite{wang2022intelligentomni, fang2022intelligent, benaya2023physical} studied the IOS aided unmanned aerial vehicle communications, whereas \cite{wang2022safeguarding} compared the impact of ES-IOS and MS-IOS on non-orthogonal multiple access. Furthermore, \cite{zhang2022dual} proposed an IOS based beam training mechanism to reduce the training-cost. The above methods, however, suffer from various challenges. Firstly, the perfect channel state information (CSI) of sub-channels (i.e., BS-UEs channel, BS-IOS channel, and IOS-UEs channel) is essential in these schemes \cite{wu2021intelligent}. Due to the passive feature, IOS fails to sense the incident signal, which renders the CSI acquisition's difficulty. Secondly, even if the accurate CSI is available, traditional optimization based solutions to IOS control often results in prohibitive computational complexity \cite{xu2023intelligent}. Thirdly, current schemes are built on the accurate IOS modeling, but it becomes impractical in dynamic wireless networks, especially when UEs move randomly. Therefore, it’s imperative to design an IOS control scheme that adapts to the channel’s time-varying feature and the UEs’ random mobility.

This work develops an online learning framework, referred to as \textit{Deep reinforcement learning IOS (DeepIOS) with digital twins}, to optimize IOS configurations on the fly. In our design, deep reinforcement learning (DRL) \cite{mnih2015human} enables \textit{model-free} optimization, whereas digital twin \cite{barricelli2019survey} guarantees \textit{real-time} decision-making.
Specifically, in DRL, a decision-making agent (e.g., the IOS controller) interacts with the external environment (e.g., the MU-MIMO system) by conducting an action, and acquires the feedback in the form of the reward that reflects how good the action was. Over the lifetime, the agent searches for a strategy that maximizes its rewards, without knowing prior environmental information \cite{sutton2018reinforcement}. DRL, however, consumes a lot of time that can be used for data transmission to perform trial-and-error interactions. Digital twin is emerging as a possible solution, which refers to physical entities' virtual representation, powered with sheer amounts of historical and fresh data related to physical entities to monitor and optimize them in real time \cite{nguyen2021digital}. In digital space, data interaction happens among virtual objectives (i.e., inter-twin communication) with negligible computational overhead and transmission time \cite{wu2021digital}. This interaction, thus, runs thousands of times (even more) in a transmission time interval of the realistic system (e.g., 1ms in cellular networks) without any effort.

Currently, there is little research work on DRL based IOS, and previous works \cite{zhao2022simultaneously, adhikary2023artificial, luo2024meta} only focused on how to optimize the phase-shift, while ignoring the amplitude configuration. In \cite{zhao2022simultaneously}, the authors investigated the joint unmanned-aerial-vehicle (UAV) trajectory and IOS phase-shift optimization problem in IOS assisted UAV networks. The sum data rate is maximized while satisfying the UAV’s flight safety, by using a novel distributed robust DRL algorithm. The authors of \cite{adhikary2023artificial} considered an IOS aided integrated sensing and communication system, where the phase-shift configuration of IOS and the transmit power of BS are jointly optimized to maximize the sensing utility function while satisfying the preset communication requirement. To recognize different network environments and automatically retrain the learning model, a meta-critic DRL framework was proposed in \cite{luo2024meta} to maximize the sum data rate of the IOS assisted MU-MIMO system. Unlike the above works, this paper jointly optimizes the phase-shift and amplitude of IOS. In addition, the DRL based IOS configuration schemes in \cite{zhao2022simultaneously, adhikary2023artificial, luo2024meta} cannot guarantee real-time decision-making, since they consume a lot of time performing online-training. Instead, by integrating DRL with digital twins, DeepIOS attains it. Several digital twins assisted wireless network architectures can be found in \cite{fan2021digital, huang2024digital, peng2024stochastic}. Specifically, in \cite{fan2021digital}, a digital twin empowered mobile edge computing architecture was proposed to guarantee real-time sensing and computing safety; in \cite{huang2024digital}, a digital twin driven video streaming was developed to enable network virtualization and tailored network management; and in \cite{peng2024stochastic}, a digital twin assisted heterogeneous network was considered to minimize the system energy consumption. Unlike existing works, we focus on the exploration of digital twins in IOS configurations. To the best of our knowledge, the integration of DRL with digital twins has not been explored in IOS assisted MU-MIMO systems, which therefore motivates this work.


Benefiting from the in-depth integration of DRL and digital twin, three closed loops are formed in the proposed framework to promote each other for efficient IOS control. To be specific, in the first closed-loop that operates in the digital space, the digital DeepIOS agent frequently interacts with the digital twin module to generate a large number of experiences and then adopts them to train the deep neural network (DNN) \cite{DeepLearning}. In the second closed loop that operates in the physical space, the physical DeepIOS agent (i.e., the IOS controller) configures IOS parameters to improve the MU-MIMO system's performance, without acquiring sub-channels' CSI and UEs’ positions. Furthermore, it only executes decisions without online training that consumes time, thereby satisfying the decision-making's real-time feature. In the third closed-loop, the digital DeepIOS agent delivers the trained DNN to the physical DeepIOS agent for policy improvement, and meanwhile the physical space provides fresh real data from the practical environment to calibrate the digital twin module.

Overall, this paper's \textit{major contributions} are below:

\textit{1) DeepIOS Scheme with Action Branch Architecture:} We design a DeepIOS scheme for IOS configurations, including phase-shift and amplitude, in MU-MIMO systems. As a first step, the sum-rate maximization problem is reformulated into a partial observable Markov decision process (POMDP) \cite{sutton2018reinforcement}, by properly defining the state, action, and reward in DRL paradigms. Thereafter, the deep Q-network (DQN) algorithm \cite{mnih2015human} implements DeepIOS, which makes configuration decisions through the $\epsilon$-greedy strategy while training the DNN through the fixed Q-target and experience replay mechanisms. To further reduce the computational complexity, we introduce an action branch architecture \cite{tavakoli2018action} for incorporation into DeepIOS. Specifically, traditional DRL, e.g., DQN, is limited to the applications with a small action space. However, multiple parameters of IOS, i.e., (phase-shift, amplitude), are required to be configured in our system, and DQN usually optimizes them in a combinatorial manner. In this regard, the optional action combination of DeepIOS is the number of optional phase-shifts multiplied by the number of optional amplitudes. Consequently, the DNN of DeepIOS needs to evaluate a lot of actions with expensive computational overhead, and undoubtedly learns ineffective policies. With the action branch architecture, DeepIOS optimizes the phase-shift and amplitude in a parallel and separate manner, thereby reducing the number of actions that the DNN needs to evaluate.

\textit{2) Real-Virtual Modeling:} We develop a real-virtual coexistence technology between the IOS empowered MU-MIMO environment and digital twins to ensure real-time decision-making for DeepIOS. The constructed framework is a closed-loop control system, where the IOS empowered MU-MIMO environment provides historical and fresh data to build a digital twin module synchronized with the real environment, whereas the digital twin serves as a pre-validation platform for IOS configurations in realistic systems. In the physical space, the IOS is treated as part of the wireless channel, while the MU-MIMO system and wireless channel are considered as a digital twin module interacting with the virtual IOS controller in the digital space. The separation of IOS from the wireless communication system enhances the independence of IOS, which speeds up its rollout in realistic systems.

\textit{3) Digital Twin Construction:} We put forth a new supervised learning approach to construct the digital twin module. In particular, how digital twin, despite its powerful capabilities, can be constructed to map the real wireless environment into the digital space remains an open question. In our work, a supervised DNN model \cite{caruana2006empirical} is trained for learning the IOS empowered MU-MIMO system's dynamics, from historical transmission data. In a sense, the digital twin module is a small generative model that derives equivalent CSI and transmission results based on the imported IOS configuration. Furthermore, with the fresh data from the realistic system, the generative model is calibrated to cater the real environment.

\textit{4) Comprehensive Empirical Evaluation:} We compare the performance of different algorithms through simulation, and demonstrate that compared with random and MAB algorithms, DeepIOS achieves a higher data rate and is more robust to different parameter settings, even without prior sub-channel's CSI and UEs' mobility. In addition, the action branch architecture greatly reduces DeepIOS's training complexity. More importantly, the digital twin module significantly improves DeepIOS's convergence speed and simulation run-time, which are two very important features in realistic systems.

\textit{Notations:} Vectors (matrices) are denoted by boldface lower (upper) case letters, $\mathbb{C}^{N{\times}M}$ represents the space of $N{\times}M$ complex matrix, $|\cdot|$ represents the absolute value, $||\cdot||_2$ represents the 2-norm, $\text{diag}\{\boldsymbol{a}\}$ represents a square diagonal matrix with the elements of vector $\boldsymbol{a}$ as diagonal elements, $\mathbb{E}[\cdot]$ represents the statistical expectation, ${\odot}$ represents the Hadamard (element-wise) product, and $(\cdot)^H$ denotes the conjugate transpose.
For ease readability, we list the major notations of this paper in TABLE \ref{tab:table1}.

\begin{table}[]
    \centering
    \caption{Major notations of this paper.}
    \renewcommand\arraystretch{1.2}
    \begin{tabular}{|c|c|}\hline
     $N$ & Number of antennas at BS \\\hline
     $K$/$\mathcal{K}$ & Number/Set of UEs \\\hline
     $M$/$\mathcal{M}$ & Number/Set of IOS elements \\\hline
     $\varphi_m$/$\Delta{\varphi}_m$ & Phase-shift/phase-shift increment of IOS element $m$ \\\hline
     ${\beta}_{\text{r},m}$/${\beta}_{\text{t},m}$ & Reflecting/Refracting amplitude of IOS element $m$ \\\hline
     ${\boldsymbol\Phi_{\text{r}}}$/${\boldsymbol\Phi_{\text{t}}}$ & Reflecting/Refracting coefficient of IOS \\\hline
     $\lambda$ & Rician factor \\\hline
     $\widetilde{\textbf{H}}(t)$ & Estimated channel matrix \\\hline
     $R_k$ & Data rate at UE $k$ \\\hline
     $Q(s, a)$ & Action-value function (Q-value) \\\hline
     $s(t)$/$s(\tau)$ & State of the agent \\\hline
     $a(t)$/$a(\tau)$ & Action of the agent \\\hline
     $r(t)$/$r(\tau)$ & Reward of the agent \\\hline
     $e(t)$/$e(\tau)$ & Esperience of the agent \\\hline
     $\pi$/${\pi}^*$ & policy/optimal policy \\\hline
     $\boldsymbol{\theta}_{\text{s}}$/$\boldsymbol{\theta}_{\text{1}}$/$\boldsymbol{\theta}_{\text{2}}$/$\boldsymbol{\theta}_{\text{p}}$/$\boldsymbol{\theta}_{\text{r}}$ & DNN parameter vector \\\hline
     $\omega$/$\gamma$/$\alpha$ & Penalty factor/Discount factor/Learning rate \\\hline
     $\epsilon_{\text{p}}$/$\epsilon_{\text{d}}$ & Exploration probability \\\hline
     $\alpha$ & Learning rate \\\hline
     $L_1$/$L_2$ & Size of the action set $\mathcal{A}_1$/$\mathcal{A}_2$ \\\hline
     $\mathcal{E}$/$E$ & Experience buffer/its size \\\hline
     $\mathcal{B}_{\text{E}}$ & Size of mini-batch \\\hline
     $N_{\text{E}}$/$N_{\text{D}}$ & Number of experiences used for training\\\hline
     $T_0$/$T_1$ & Update frequency of the target DNN/digital twins\\\hline
     $\Gamma$ & Number of interactions in the digital space \\\hline
     $\mathcal{D}$/$D$ & Data requisition module/its size \\\hline
    \end{tabular}
    \label{tab:table1}
\end{table}

\section{System Model}\label{1}

\begin{figure}[t]
	\centering
	\includegraphics[scale=0.50]{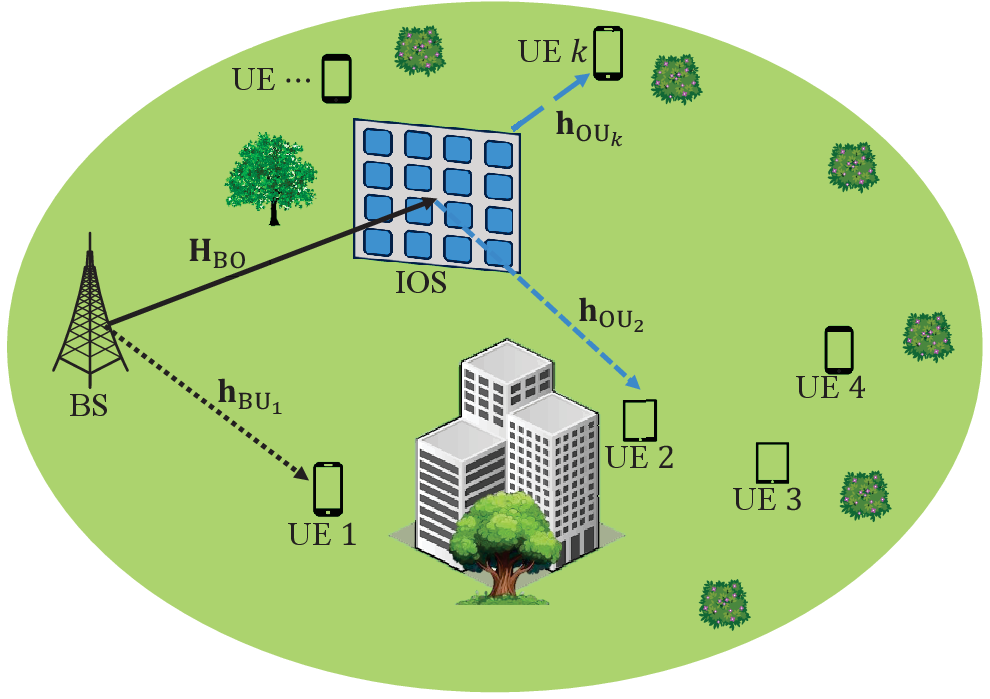}
	\caption{IOS assisted MU-MIMO systems. }
	\label{system}
\end{figure}

As shown in Fig. \ref{system}, we consider a downlink MU-MIMO system, where a BS equipped with $N$-antenna serves $K$ single-antenna mobile UEs that are denoted by $\mathcal{K}=\{1,\,2,\,\cdots,\,K\}$. Due to complex scattering characteristics and various blockages, some UEs that are far from the BS undergo severe fading, resulting in the low quality of service. To improve the receipt signals' strength at UEs, an IOS is deployed in the system.

\subsection{IOS Model}
The IOS is composed of $M$ reconfiguration elements that are denoted by $\mathcal{M}=\{1,\,2,\,\cdots,\,M\}$. The reflected and refracted signals' phase-shifts can be either the same or different, which hinge on the IOS elements' structure \cite{zhang2022intelligent1}. This system considers the case where the reflected and refracted signals share the same phase-shift. Hence, the phase-shift vector of IOS is $\boldsymbol{\varphi}=[{\varphi}_1,\,{\varphi}_2,\,\cdots,\,{\varphi}_M]$, where ${\varphi}_m$ is the quantized phase picked from the set $\mathcal{B}=\{-\pi,\,-\pi+\pi/{2^{r-1}},\,-\pi+\pi/{2^{r-2}},\,\cdots,\,\pi\}$ with $2^{r}+1$ possible values. Furthermore, the configurations of IOS under different protocols are different, and the details are given as follows.

\textit{1) ES Protocol:} all elements are operated in the simultaneous transmission and reflection mode, each of which splits the incident signal into reflected and refracted signals with different amplitudes. Let ${\beta}_{\text{r},m}$ and ${\beta}_{\text{t},m}$ denote the reflecting and refracting amplitudes of element $m$, respectively. Thus, the constraints for the ES-IOS can be given by
\begin{align}
{\beta}_{\text{r},m}^2+{\beta}_{\text{t},m}^2=1,\,\,{\beta}_{\text{r},m}\,{\in}\,[0,1],\,\,{\beta}_{\text{t},m}\,{\in}\,[0,1],\,\,\forall\,m{\in}{\mathcal{M}}. \label{E1}
\end{align}

\textit{2) MS Protocol:} all elements are partitioned into two groups: one group is operated in the reflection mode, while the other group is operated in the refraction mode. Thus, the constraints for the MS-IOS can be given by
\begin{align}
{{\beta}_{\text{r},m}}+{{\beta}_{\text{t},m}}=1,\,\,{\beta}_{\text{t},m} {\in} \{0,\,1\}\,\,{\beta}_{\text{r},m} {\in} \{0,\,1\},\,\,\forall\,m{\in}{\mathcal{M}}, \label{E2}
\end{align}
where element $m$ under MS-IOS operates in either reflection or refraction mode.

Thus, the reflecting and refracting amplitude and phase-shift coefficient matrices of IOS (ES-IOS/MS-IOS) are
\begin{subequations}
\begin{align}
{\boldsymbol\Phi_{\text{r}}}=\text{diag}\{{\beta}_{\text{r},1}{e^{j{\varphi}_1}},\,{\beta}_{\text{r},2}{e^{j{\varphi}_2}},\,\cdots,\,{\beta}_{\text{r},M}{e^{j{\varphi}_M}}\},\label{E3a}\\
{\boldsymbol\Phi_{\text{t}}}=\text{diag}\{{\beta}_{\text{t},1}{e^{j{\varphi}_1}},\,{\beta}_{\text{t},2}{e^{j{\varphi}_2}},\,\cdots,\,{\beta}_{\text{t},M}{e^{j{\varphi}_M}}\}.\label{E3b}
\end{align}
\end{subequations}



\subsection{Channel Model}
Let ${\textbf{h}}_{\text{B}\text{U}_k}\,{\in}\,{\mathbb{C}}^{N{\times}1}$ denote the channel response vector between BS and UE $k{\in}\mathcal{K}$, ${\textbf{H}}_{\text{B}\text{O}}\,{\in}\,{\mathbb{C}}^{N{\times}M}$ denote the channel response matrix between BS and IOS, and ${\textbf{h}}_{\text{O}\text{U}_k}\,{\in}\,{\mathbb{C}}^{M{\times}1}$ denote the channel response vector between UE $k$ and IOS. Thus, the channel model between BS and UE $k$ is ${\textbf{h}}_{k}={\textbf{h}}_{\text{B}\text{U}_k}+{\textbf{H}}_{\text{B}\text{O}}{\boldsymbol\Phi_{i}}{\textbf{h}}_{\text{O}\text{U}_k}$,
wherein $i$ is $\text{r}$ if UE $k$ is the reflected UE, otherwise $i$ is $\text{t}$. Furthermore, the multi-UE channel, i.e., the aggregated equivalent channel, is given by ${\textbf{H}}={\textbf{H}}_{\text{B}\text{U}}+{\textbf{H}}_{\text{B}\text{O}}{\boldsymbol\Phi_{i}}{\textbf{H}}_{\text{O}\text{U}}$,
where ${\textbf{H}}=[{\textbf{h}}_{1},\,{\textbf{h}}_{2},\,\cdots,\,{\textbf{h}}_{K}]$, ${\textbf{H}}_{\text{B}\text{U}}=[{\textbf{h}}_{\text{B}\text{U}_1},\,{\textbf{h}}_{\text{B}\text{U}_2},\,\cdots,$ ${\textbf{h}}_{\text{B}\text{U}_K}]$, and ${\textbf{H}}_{\text{O}\text{U}}=[{\textbf{h}}_{\text{O}\text{U}_1},\,{\textbf{h}}_{\text{O}\text{U}_2},\,\cdots,\,{\textbf{h}}_{\text{O}\text{U}_K}]$.

The Rician channel model is employed for ${\textbf{h}}_{\text{B}\text{U}_k}$, ${\textbf{H}}_{\text{B}\text{O}}$, and ${\textbf{h}}_{\text{O}\text{U}_k}$. Taking ${\textbf{H}}_{\text{B}\text{O}}$ as an example, it can be represented as
\begin{equation}
    {\textbf{H}}_{\text{B}\text{O}}=\sqrt{\frac{\lambda}{\lambda+1}}{\textbf{H}}_{\text{B}\text{O}}^{\text{LOS}}+\sqrt{\frac{1}{\lambda+1}}{\textbf{H}}_{\text{B}\text{O}}^{\text{NLOS}},
\end{equation}
where $\lambda$ is the Rician factor, ${\textbf{H}}_{\text{B}\text{O}}^{\text{LOS}}$ is the position-dependent line-of-sight (LoS) component, and ${\textbf{H}}_{\text{B}\text{O}}^{\text{NLOS}}$ is the non-LoS (NLoS) component with fast-fading, which accounts for independent and identically distributed (i.i.d) complex circularly-symmetric random variables following $\mathcal{C}\mathcal{N}(0, 1)$. We consider a time-slotted system, i.e., $t=0,\,1,\,2,\cdots$, where the NLoS component varies over time slots and it is i.i.d in different time slots. Furthermore, each UE moves randomly in each time slot, and thus the LoS component, being position-dependent, also changes in each time slot.

\subsection{Transmission Model}
\begin{figure}[t]
	\centering
	\includegraphics[scale=0.42]{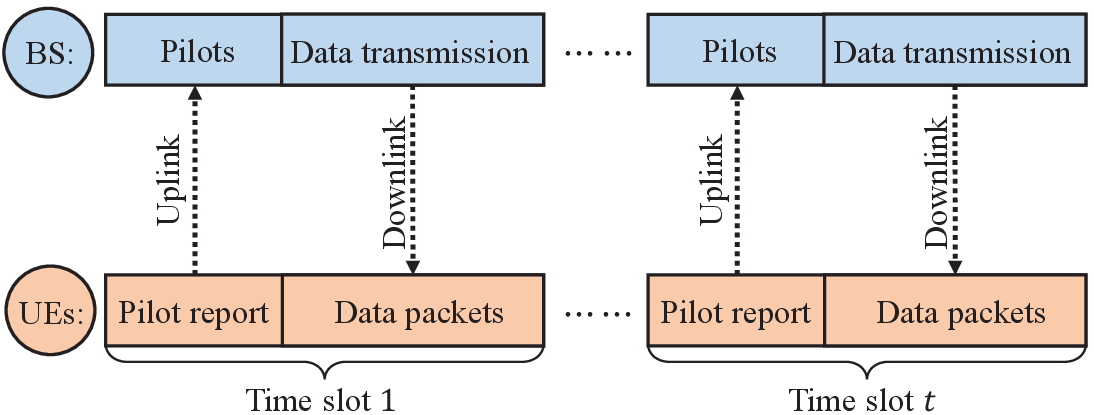}
	\caption{Operations within a time slot. }
	\label{TimeSlot}
\end{figure}

The system assumes the time-division duplexing, and thus the BS estimates the downlink channel quality from the uplink channels' pilots based on the channel reciprocity \cite{wang2022intelligent}. As shown in Fig. \ref{TimeSlot}, in each time slot, there are two stages: (i) uplink pilot report, and (ii) downlink data transmission.

\textit{1) Uplink Pilot Report:} all UEs transmit pilots to BS at the same time, and thus the received pilot signal is ${\textbf{Y}}_{\text{p}}={\textbf{H}}{\textbf{X}}_{\text{p}}+{\textbf{N}}$,
where ${\textbf{X}}_{\text{p}}\,{\in}\,\mathbb{C}^{K{\times}K}$ is the pilot matrix, and ${\textbf{N}}\,{\in}\,\mathbb{C}^{N{\times}K}$ is the additive white Gaussian noise. With ${\textbf{Y}}_{\text{p}}$, the BS estimates the channel matrix $\widetilde{\textbf{H}}$ via the minimum mean square error, i.e., ${\widetilde{\textbf{H}}}={\textbf{Y}}_{\text{p}}{\textbf{X}}_{\text{p}}^{H}{\big(}{\textbf{X}}_{\text{p}}{\textbf{X}}_{\text{p}}^{H}+{{\sigma}_{\text{p}}^2}{\textbf{I}}{\big)}^{-1}$,
where ${{\sigma}_{\text{p}}^2}$ is the noise variance at BS, ${\textbf{I}}$ is the identity matrix, and $(\cdot)^H$ is the matrix transpose.

\textit{2) Downlink Data Transmission:} by using zero-forcing\footnote{Owing to the low complexity, this paper uses zero-forcing for precoding at BS. It's worth noting that other precoding approaches can also be employed.}, the precoding matrix for downlink transmission is ${\textbf{V}}=[$ ${\textbf{v}}_1,$ ${\textbf{v}}_2,\,\cdots,\,{\textbf{v}}_K]^H={\textbf{U}}{\big(}{{\widetilde{\textbf{H}}}^H}{\widetilde{\textbf{H}}}{\big)}^{-1}{{\widetilde{\textbf{H}}}^H}$,
where ${\textbf{U}}=\text{diag}\{1/{\Vert {\textbf{v}_1} \Vert}_2,$ $1/{\Vert {\textbf{v}_2} \Vert}_2,\,\cdots,\,1/{\Vert {\textbf{v}_K} \Vert}_2\}$ is the normalized power. Thus, the signal-to-interference-and-noise (SINR) ratio at UE $k$ is
\begin{equation}
    {\text{SINR}_k}=\frac{|{\textbf{v}_k^H}{\textbf{h}_k}|^2}{\sum\limits_{l \neq k}^{K}|{\textbf{v}_l^H}{\textbf{h}_k}|^2+{{\sigma}_k^2}},
\end{equation}
where ${{\sigma}_k^2}$ is the noise variance at UE $k$. Furthermore, the transmission data rate at UE $k$ is
\begin{equation}
    {R_k}={\text{log}_2}(1+{\text{SINR}_k}).\label{datarate}
\end{equation}

\subsection{Problem Formulation}\label{ProblemFormulation}
We aim to maximize the system's sum data rate, by optimizing the reflecting and refracting coefficient matrices of IOS, i.e., ${\boldsymbol\Phi_{\text{r}}}$ and ${\boldsymbol\Phi_{\text{t}}}$. The corresponding formulation is then:
\begin{subequations}
\begin{align}
\max\limits_{{\boldsymbol\Phi_{\text{r}}}, {\boldsymbol\Phi_{\text{t}}}}\,\,&\sum\limits_{k=1}^{K}{R_k} \label{ProblemFormulation1}\\
\text{s.t.}\,\,&\eqref{E1}\,\,\text{or}\,\,\eqref{E2}, \label{ProblemFormulation2}\\
&{\varphi}_m{\in}{\mathcal{B}},\,\,{m}{\in}{\mathcal{M}}. \label{ProblemFormulation3}
\end{align}
\end{subequations}

In constraint \eqref{ProblemFormulation2}, the first item is for the ES-IOS case, whilst the second item is for the MS-IOS case. Constraint \eqref{ProblemFormulation3} provides the set of optional phase-shifts.
\textit{The traditional optimization approach cannot address the formulated problem} in that (i) this is a combinatorial optimization problem, which is usually NP-hard; (ii) it is difficult to obtain accurate sub-channels' CSI, i.e., ${\textbf{H}}_{\text{B}\text{U}}$, ${\textbf{H}}_{\text{B}\text{O}}$, and ${\textbf{H}}_{\text{O}\text{U}}$, due to the time-varying characteristic of the wireless channel and the passive feature of the IOS; and (iii) all UEs' locations change in real time. Therefore, this paper employs the DRL technique to tackle these challenges.

\section{DeepIOS Scheme}\label{2}
This section first recasts the formulated problem as a POMDP. After that, the DQN algorithm implements DeepIOS. Finally, to reduce the computational complexity, DeepIOS introduces the action branch architecture.

\subsection{POMDP Formulation}\label{POMDPFormulation}
The formulated IOS reflecting and refracting coefficient matrices optimization problem can be characterized as a POMDP \cite{sutton2018reinforcement}, which is described by a six-tuple $(\mathcal{O}, \mathcal{S}, \mathcal{A}, \mathcal{P}, \mathcal{R}, \gamma)$.

\textit{1) Agent and Environment $\mathcal{O}$:} the IOS and its controller form a DeepIOS agent, where the IOS is the agent's body, and its controller is the agent's brain that guides the behavior of IOS. The environment is composed of the complete MU-MIMO system, i.e., BS, wireless channel, and UEs. In our design, the operations of the MU-MIMO system and the IOS are stand-alone: the MU-MIMO system is almost unaware of IOS's existence, other than the need to feed its instantaneous performance back to the IOS controller; and the IOS is also unaware of the MU-MIMO system's working mechanism. The IOS with DeepIOS, hence, can be deployed in various communication systems, e.g., cellular and WLAN, without redesigning their original protocols.

\textit{2) Observable State Space $\mathcal{S}$:} the agent's observable information is employed as the state $s(t)$. Specifically, apart from the estimated channel matrix $\widetilde{\textbf{H}}(t)$ in the current time slot, the reflecting and refracting coefficient matrices, i.e., $\boldsymbol{\Phi}_{\text{r}}(t-1)$ and $\boldsymbol{\Phi}_{\text{t}}(t-1)$, from the previous time slot can also be acquired. We convert $\widetilde{\textbf{H}}(t)\,{\in}\,{\mathbb{C}^{N{\times}K}}$ into a $2{\times}N{\times}K$ real-valued tensor with real and imaginary parts stored separately. Likewise, $\boldsymbol{\Phi}_{\text{r}}(t-1)\,{\in}\,{\mathbb{C}^{M{\times}M}}$ and $\boldsymbol{\Phi}_{\text{t}}(t-1)\,{\in}\,{\mathbb{C}^{M{\times}M}}$ can also be converted into two $2{\times}M{\times}M$ real-valued tensors. The state $s(t)$, thus, consists of three parts: (i) the estimated channel matrix $\widetilde{\textbf{H}}(t)$ with dimension $2{\times}N{\times}K$, (ii) the reflecting coefficient matrix $\boldsymbol{\Phi}_{\text{r}}(t-1)$ with dimension $2{\times}M{\times}M$, and (iii) the refracting coefficient matrix $\boldsymbol{\Phi}_{\text{t}}(t-1)$ with dimension $2{\times}M{\times}M$. The state space $\mathcal{S}$ consists of all possible $s(t)$.


\textit{3) Action Space $\mathcal{A}$:} the action space is composed of all feasible actions that can be executed by the agent. Since the agent aims to optimize the reflecting and refracting coefficient matrices $\boldsymbol{\Phi}_{\text{r}}$ and $\boldsymbol{\Phi}_{\text{t}}$, its action $a(t)$ in each time slot is an appropriate pair ${\big(}\boldsymbol{\Phi}_{\text{r}}(t), \boldsymbol{\Phi}_{\text{t}}(t){\big)}$. According to \eqref{E1} and \eqref{E2}, once the reflection amplitude is selected, the refraction amplitude is determined accordingly. Furthermore, the reflected and refracted signals share the same phase-shift. Thus, the agent only needs to focus on $\boldsymbol{\Phi}_{\text{r}}$. To enhance the decision's efficiency, the agent adjusts the phase-shift's increment $\Delta{\varphi}(t)$ in time slot $t$, instead of directly selecting a new phase-shift. In other words, the agent's sub-action in phase-shift selection is $\Delta{\varphi}(t)$, and its phase-shift is updated by
\begin{equation}
    {\varphi}_m(t+1)={\varphi}_m(t){\odot}\Delta{\varphi}(t),
\end{equation}
where $\odot$ is the Hadamard (element-wise) product, and the phase-shifts $\{{\varphi}_m(t)|m{\in}{\mathcal{M}}\}$ are all initialized to 1.
We employ the discrete Fourier transform vectors \cite{wang2022intelligent} as the optional $\Delta{\varphi}(t)$ set $\mathcal{A}_1$, which could be ${\big{\{}}\textbf{w}(-L_1/M),\cdots,\textbf{w}(-1/M),$ $\textbf{w}(0),\textbf{w}(1/M),\cdots,\textbf{w}(L_1/M){\big{\}}}$ if its size is $2{L_1}+1$. Without loss of generality, both reflective and refractive arrays are assumed to be uniform linear arrays. Thus, ${\textbf{w}(\phi)}$ is denoted by ${\big{[}}1,{e^{j{\pi}{\phi}}},\cdots,{e^{j(M-1){\pi}{\phi}}}{\big{]}}^{T}$,
where $\phi=-l/M$ and $l/M$ $(l=1,\,2,\,\cdots,\,L_1)$ are towards the opposite directions, which enables the agent to quickly correct a negative action; and a large $l$ can speed up the transition among phase-shifts. With the above design, the agent's first sub-action is ${a_1}(t)=\Delta{\varphi}(t)$.

On the other hand, the agent's second sub-action ${a_2}(t)$ is the amplitude ${\beta}_{\text{r},m}(t)$, and the corresponding sub-action set is $\mathcal{A}_2$. Under the ES-IOS protocol, we consider the equal mode \cite{cai2022joint}, where all elements are equipped with the same amplitude, i.e., ${\beta}_{\text{r},m}={\beta}_{\text{r}}$, and the number of optional ${\beta}_{\text{r}}$ is given as $L_2$; under the MS-IOS protocol, we also consider $L_2$ types of different element groups, where ${\beta}_{\text{r},m}, \forall m{\in}{\mathcal{M}}$ in a specific element group is 0 or 1 in random.

As in \eqref{E3a}, the amplitude and phase-shift are presented in a combinatorial manner, i.e., $a(t)={\big{(}}{a_1}(t), {a_2}(t){\big{)}}$. Thus, the total action space's size is $(2{L_1}+1){\times}{L_2}$, which indicates that the DNN needs to evaluate $(2{L_1}+1){\times}{L_2}$ actions during execution and training. In Section \ref{DeepIOSImplementation}, the action branch architecture is introduced to save this computational overhead by reducing $(2{L_1}+1){\times}{L_2}$ to $2{L_1}+1+{L_2}$.

\textit{4) Transition Probabilities $\mathcal{P}$:} after taking action $a(t)$, the probability of the state transition from $s(t)$ to $s(t+1)$ is denoted by ${\mathcal{P}}{\big{(}}s(t+1)|s(t),a(t){\big{)}}$. Due to UEs' random mobility and channels' random fading, the underlying transition probabilities among different observable states are non-stationary. Our designed system consequently is a POMDP, where $s$ is the state containing the partial observation from the MU-MIMO system and $a$ is determined under the uncertainty. These transition probabilities $\mathcal{P}$ are unknown since they are determined by the unknown UEs' position information and sub-channels' CSI, which jointly reflect the MU-MIMO system's dynamics and randomness.

\textit{5) Reward Function $\mathcal{R}$:} the reward $r(t+1)$, provided by the environment, reflects how well the action $a(t)$ was performed under the state $s(t)$ in time slot $t$. Thus, $r(t+1)$ can be expressed as $\mathcal{R}{\big{(}}s(t),a(t){\big{)}}$. Since the system objective is to maximize the sum data rate, the most straightforward design of the reward function is to employ the sum data rate as the reward of the agent. However, under this design, the agent cannot explore a sufficiently efficient IOS configuration strategy. According to \cite{wang2022intelligent}, to enable the agent to maximize the system performance, a preset data rate threshold $R_{\text{th}}$ can be introduced into $r(t+1)$. Thus, the reward is denoted by
\begin{equation}
    r(t+1)= \begin{cases}
\sum\limits_{k=1}^{K}{R_k}(t),\quad &\text{if}\,\,\sum\limits_{k=1}^{K}{R_k}(t) \geq R_{\text{th}}, \\
\sum\limits_{k=1}^{K}{R_k}(t)-\omega,\quad & \text{otherwise},
\end{cases}\label{rewardfunction}
\end{equation}
where $\omega$ is the penalty factor. Specifically, after receiving the data packet, all UEs feedback their obtained data rate to the agent. The sum data rate is adopted as the reward if this value is higher than $R_{\text{th}}$, indicating that the action $a(t)$ maintains an acceptable performance; otherwise, a penalty factor $\omega$ is added to encourage the agent to improve its policy.


\textit{6) Discount Factor $\gamma$:} the discount factor $\gamma{\in}[0,1]$ determines the importance of current and future rewards. With $\gamma$, the cumulative discounted reward is defined as ${G(t)} =\sum_{l = t}^\infty  {{\gamma ^{l-t}}} {r(l+1)}$.
Given $s(t)$, the policy of selecting $a(t)$ can be denoted by ${\pi}{\big(}a(t)|s(t){\big)}$. In DRL, the agent aims to find an optimal policy ${\pi}^*$ that maximizes $G(t)$. Furthermore, for each state-action pair $({s,a})$, the action-value function (i.e., Q-value) under policy $\pi$ is defined as ${Q_\pi }{\big(}s(t),a(t){\big)} ={\mathbb{E}_{{s(t+1)}, {a(t+1)}, \cdots} }{\big{[}} {{G(t)}|{s(t)},{a(t)}} {\big{]}}$.
In other words, the Q-value is determined by the policy $\pi$ (as $a(t){\sim}{\pi}{\big(}a(t)|s(t){\big)}$) and the transition probabilities $\mathcal{P}$ (as $s(t+1){\sim}{\mathcal{P}}{\big{(}}s(t+1)|s(t),a(t){\big{)}}$).

Thus far, the IOS coefficient matrices optimization problem has been reformulated into the POMDP framework.

\subsection{DeepIOS Implementation}\label{DeepIOSImplementation}
\begin{figure}[t]
	\centering
	\includegraphics[scale=0.53]{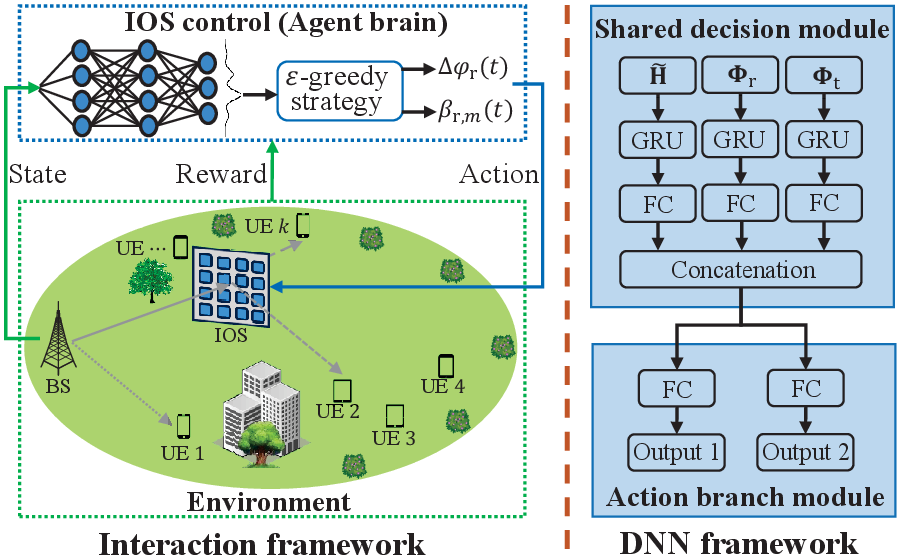}
	\caption{DeepIOS module. }
	\label{DeepIOS2}
\end{figure}

Based on the formulated POMDP problem, the DQN algorithm \cite{mnih2015human} is employed to implement the DeepIOS scheme. Afterward, the action branch architecture \cite{tavakoli2018action} is incorporated into DeepIOS for complexity reduction. Fig. \ref{DeepIOS2} presents the complete DeepIOS module, and its details are below.

\textit{1) $\epsilon$-greedy Strategy for Execution:} given the state $s(t)$, the agent approximates the Q-values for all actions, i.e., ${\big\{}Q{\big(}$ $s(t),a; \boldsymbol{\theta} {\big)}|a{\in}{\mathcal{A}}{\big\}}$, through the DNN with parameter vector $\boldsymbol{\theta}$. In general, the optimal strategy ${\pi}^*$ is to choose the action that yields the largest $Q{\big(}{s(t),a; \boldsymbol{\theta} }{\big)}$, if $Q{\big(}{s(t),a; \boldsymbol{\theta} }{\big)}$ is perfectly equivalent to ${Q_\pi }{\big(}{s(t),a}{\big)}$. However, the agent may fall into sub-optimal policies if it always greedily chooses the optimal action \cite{sutton2018reinforcement}. As a result, random exploration with a certain probability is required. The $\epsilon$-greedy strategy \cite{sutton2018reinforcement} fits exactly the above idea, which is given by
\begin{equation}\label{actionselection}
    {\pi}= \begin{cases}
{{\pi}^*}(a|s),\quad &\text{with prob.}\,\,1-\epsilon, \\
\text{random selection},\quad & \text{with prob.}\,\,\epsilon,
\end{cases} 
\end{equation}
where $\epsilon$ is the exploration probability. Specifically, in each time slot, a random action is chosen from $\mathcal{A}$ with probability $\epsilon$, whereas an action with the largest Q-value is selected with probability $1-\epsilon$.

\textit{2) Fixed Q-target Method for Training:} two DNNs, referred to as evaluation DNN and target DNN, are employed to enhance DeepIOS's stability \cite{mnih2015human}. Specifically, ${Q({{s(i)},{a(i)};\boldsymbol{\theta} })}$ is the output of the evaluation DNN with parameter vector $\boldsymbol{\theta}$; and $Q({s(i+1)},a';\boldsymbol{\theta}^-)$ is the output of the target DNN, which has the same structure as the evaluation DNN but with different parameter vector $\boldsymbol{\theta}^-$. Furthermore, these two DNNs possess different update frequencies, where $\boldsymbol{\theta}$ is trained in each time slot, whilst $\boldsymbol{\theta}^-$ is updated by $\boldsymbol{\theta}^- \leftarrow \boldsymbol{\theta}$ every $T_0$ time slots.

\textit{3) Experience Replay Mechanism for Training:} a first-in-first-out (FIFO) experience buffer $\mathcal{E}$ that stores $E$ experiences at most is used \cite{mnih2015human}. In each time slot, the DeepIOS agent stores the experience $e(t)={\big(}s(t), a(t), r(t+1), s(t+1){\big)}$ into $\mathcal{E}$. After storing enough experiences, a mini-batch $\mathcal{B}_{\text{E}}$, composed of $N_{\text{E}}$ experiences that are sampled from $\mathcal{E}$, is employed to calculate the following loss function:
\begin{align}\label{loss}
	Loss(\boldsymbol{\theta} ) = \frac{1}{N_{\text{E}}}{\sum\limits_{e(i) \in \mathcal{B}_{\text{E}}} {{\Big(}{{y(i+1) - Q{\big(}{{s(i)},{a(i)};\boldsymbol{\theta} }{\big)}}}{\Big)}} ^2},
\end{align}
where ${y(i+1)} = {r(i+1)} + \gamma \mathop {\max }\nolimits_{a'} Q{\big(}{{s(i+1)},a';\boldsymbol{\theta}^-}{\big)}$ is the target value. Finally, the loss function \eqref{loss} is minimized to train $\boldsymbol{\theta}$ through the stochastic gradient descent (SGD) \cite{sutton2018reinforcement}.

\textit{4) Action Branch Architecture for DNN Implementation:} as mentioned in Section \ref{POMDPFormulation}, the decision-making of multiple sub-actions in a combinatorial form, i.e., $a(t)={\big{(}}{a_1}(t),$ ${a_2}(t){\big{)}}$, greatly increases the total number of actions that the DNN needs to evaluate \cite{qian2020noma}. As a consequence, original DQN falls into an inefficient policies. To alleviate this problem, we design the action branch architecture for incorporate into DeepIOS.
As illustrated in Fig. \ref{DeepIOS2}, DeepIOS's complete DNN architecture consists of two major modules: (i) the shared decision module for maintaining the connection among different sub-actions, and (ii) the action branch module for allocating an independent branch to each sub-action set. The details of the two modules are provided below.

Shared Decision Module: this module is composed of three input layers, three gated unite recurrent (GRU) layers \cite{cho2014learning}, three fully connected (FC) layers \cite{DeepLearning}, and one concatenation layer \cite{DeepLearning}. At the beginning of each time slot, the input layers first pre-process the state $s(t)$ including $\widetilde{\textbf{H}}(t)$, $\boldsymbol{\Phi}_{\text{r}}(t-1)$, and $\boldsymbol{\Phi}_{\text{t}}(t-1)$. Afterward, each GRU layer extracts the underlying temporal correlation property from the input sequence;\footnote{Our previous work \cite{ye2021multi} has demonstrated GRU's powerful capability to reason the complex temporal correlation in wireless communication systems.} subsequently, the collected feature is imported into the FC layer for analysis. Finally, a concatenation layer aggregates useful features extracted from different components, i.e., $\widetilde{\textbf{H}}(t)$, $\boldsymbol{\Phi}_{\text{r}}(t-1)$, and $\boldsymbol{\Phi}_{\text{t}}(t-1)$, then outputs them to compute different sub-actions' Q-values on action branches.

Action Branch Module: to maintain independence between two different sub-actions $a_1$ and $a_2$, a separate action branch consisting of one FC layer and one output layer is assigned to each sub-action set. The number of output ports (i.e., Q-values) on each action branch corresponds to the size of each sub-action set. Suppose that the action branch 1 and 2 evaluate the Q-values of ${a_1}{\in}{\mathcal{A}_1}$ and ${a_2}{\in}{\mathcal{A}_2}$, respectively. Thus, on action branch 1, the Q-value set is ${\big\{}Q{\big(}s(t),{a_1};{\boldsymbol{\theta}_{\text{s}}},{\boldsymbol{\theta}_1}{\big)}|{a_1}$ ${\in}{\mathcal{A}_1}{\big\}}$, whilst the Q-value set on action branch 2 is ${\big\{}Q{\big(}s(t),{a_2};{\boldsymbol{\theta}_{\text{s}}},$ ${\boldsymbol{\theta}_2}{\big)}|{a_2}{\in}{\mathcal{A}_2}{\big\}}$, where ${\boldsymbol{\theta}_{\text{s}}}$, ${\boldsymbol{\theta}_1}$, and ${\boldsymbol{\theta}_2}$ are the parameter vectors of the shared decision module, action branch 1, and action branch 2, respectively. In each time slot, the shared decision module's output is delivered into various action branches, such that different sub-actions' Q-values can be generated.

Execution and Training Enhancements: according to the $\epsilon$-greedy strategy as in \eqref{actionselection}, the agent decides ${a_1}(t)$ and ${a_2}(t)$ from ${\mathcal{A}}_1$ and ${\mathcal{A}}_2$, respectively. The agent then performs the action $a(t)={\big{(}}{a_1}(t), {a_2}(t){\big{)}}$ to reflect and refract the incident signal. Furthermore, for the evaluation DNN training, the loss function in \eqref{loss} is modified as
\begin{align}\nonumber
	&Loss({\boldsymbol{\theta}_{\text{s}}},{\boldsymbol{\theta}_1},{\boldsymbol{\theta}_2} ) = \frac{1}{{{N_{\text{E}}}}}\sum\limits_{e(i) \in \mathcal{B}_{\text{E}}} {\Big(}{r(i+1)} + \gamma \mathop {\max }\limits_{{a_1}'}Q{\big(}{s(i+1)},\\
 &{a_1}';{\boldsymbol{\theta}_{\text{s}}^-},{\boldsymbol{\theta}_1^-}{\big)} - Q{\big(}{{s(i)},{a_1}(i);{\boldsymbol{\theta}_{\text{s}}},{\boldsymbol{\theta}_1} }{\big)}+ {r(i+1)} + \gamma \mathop {\max }\limits_{{a_2}'} \nonumber\\
 &Q{\big(}{{s(i+1)},{a_2}';{\boldsymbol{\theta}_{\text{s}}^-},{\boldsymbol{\theta}_2^-}}{\big)} - Q{\big(}{{s(i)},{a_2}(i);{\boldsymbol{\theta}_{\text{s}}},{\boldsymbol{\theta}_2} }{\big)}{\Big)}^2,\label{loss1}
\end{align}
where ${\boldsymbol{\theta}_{\text{s}}^-}$, ${\boldsymbol{\theta}_1^-}$, and ${\boldsymbol{\theta}_2^-}$ are the target DNN's parameter vectors. Specifically, the first line in \eqref{loss1} is the contribution of action branch 1 to the loss function, whereas the second line is the contribution from action branch 2. With SGD, the parameter vectors ${\boldsymbol{\theta}_{\text{s}}},{\boldsymbol{\theta}_1},{\boldsymbol{\theta}_2}$ are updated by
\begin{align}\nonumber
	&{\boldsymbol{\theta}_{\text{s}}}\,{\leftarrow}\,{\boldsymbol{\theta}_{\text{s}}} - {\alpha}{\nabla}Loss({\boldsymbol{\theta}_{\text{s}}},{\boldsymbol{\theta}_1},{\boldsymbol{\theta}_2} ),\,\,\, {\boldsymbol{\theta}_1}\,{\leftarrow}\,{\boldsymbol{\theta}_1} - \\
 &{\alpha}{\nabla}Loss({\boldsymbol{\theta}_{\text{s}}},{\boldsymbol{\theta}_1},{\boldsymbol{\theta}_2} ),\,\,\,{\boldsymbol{\theta}_2}\,{\leftarrow}\,{\boldsymbol{\theta}_2} - {\alpha}{\nabla}Loss({\boldsymbol{\theta}_{\text{s}}},{\boldsymbol{\theta}_1},{\boldsymbol{\theta}_2} ),\label{parameterupdate}
\end{align}
where $\alpha$ is the learning rate of the agent.




As shown in Fig. \ref{DeepIOS2}, with the action branch architecture, the total number of actions that need to be estimated by DNN is $2{L_1}+1+{L_2}$, as opposed to $(2{L_1}+1){\times}{L_2}$. Thus, as the number of sub-actions increases, the total number of actions that the DNN needs to estimate increases linearly. In this regard, DeepIOS's computational complexity can be significantly reduced, which promotes the DNN to effectively explore different actions in a given state.


\section{Digital Twin Enhanced DeepIOS}\label{3}
\begin{figure}[t]
	\centering
	\includegraphics[scale=0.36]{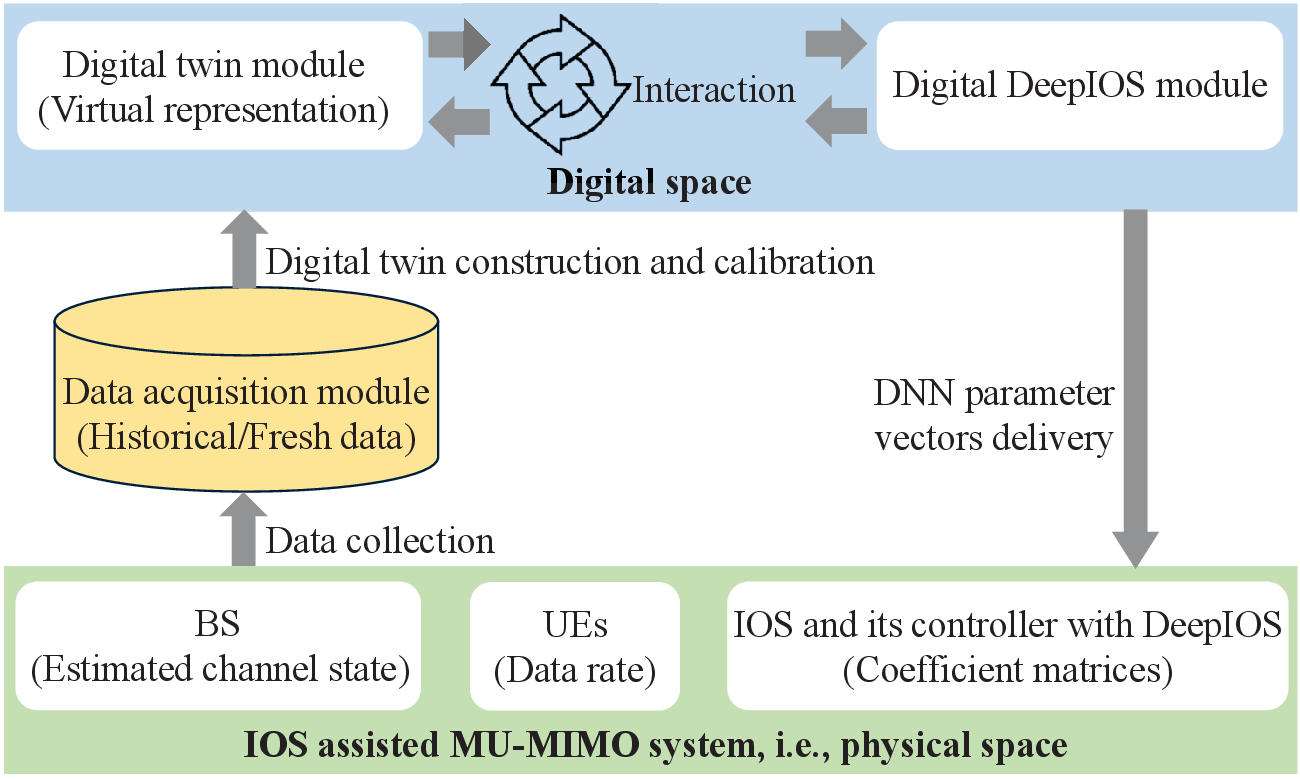}
	\caption{DeepIOS with digital twins. }
	\label{DTDeepIOS}
\end{figure}

In Section \ref{2}, we designed DeepIOS for efficient IOS configurations in MU-MIMO systems. Due to online learning, DeepIOS possesses good adaptability, but consumes a lot of time that can be used for data transmission to perform trial-and-error interactions.
To improve the decision-making's real-time, this section further introduces a digital twin enhanced DeepIOS framework that consists of both physical and digital spaces. The physical space contains all real entities, i.e., the MU-MIMO system, the IOS, the IOS controller with the DeepIOS module, and a data requisition module; whilst the digital space comprises a model-free DeepIOS module and a model based digital twin module composed of all physical entities' virtual representation. In the digital space, due to abundant computing resources, virtual objects may interact thousands of times (even more) within a time slot of the physical space \cite{mihai2022digital}. Thus, the time slot in the digital space is on a smaller time scale than that in the physical space, which is defined as ${\tau}=1,2,\cdots,\Gamma$ within a time slot $t$. Fig. \ref{DTDeepIOS} provides the complete interaction diagram between digital space and physical space, and the details are below.

\subsection{Data Requisition Module}
To build a digital twin module, large volumes of historical data that reflects the state of physical entities is essential \cite{mihai2022digital}. As shown in Fig. \ref{DTDeepIOS}, the data requisition module undertakes this task: it provides past data to support the digital twin's construction during the initialization stage, and thereafter uploads fresh data for calibration and synchronization with the real MU-MIMO system. The data collected by the data requisition module includes: (i) the estimated channel matrix from the BS; (ii) the parameter configuration from IOS, e.g., phase-shift and amplitude; and (iii) the performance matrices from UEs, e.g., data rates. Thus, the dataset in the data requisition module is
\begin{align}\nonumber
    \mathcal{D}=&{\big{\{}}c(i)={\big(}\widetilde{\text{H}}(i),{\Delta}{{\varphi}}(i),{\beta_{\text{r},1}}(i),{\beta_{\text{r},2}}(i),\cdots,{\beta_{\text{r},M}}(i),\\&{R_1}(i),{R_2}(i),\cdots,{R_K}(i){\big)}|
    i=0,1,\cdots,D-1{\big{\}}},\label{trainingdata}
\end{align}
where $D$ is the data requisition module's size. During initialization, the data requisition module is filled to build the digital twin module; after that, it updates data on an FIFO basis to calibrate the constructed digital twin module.

\subsection{Digital Twin Module}
\begin{figure}[t]
	\centering
	\includegraphics[scale=0.46]{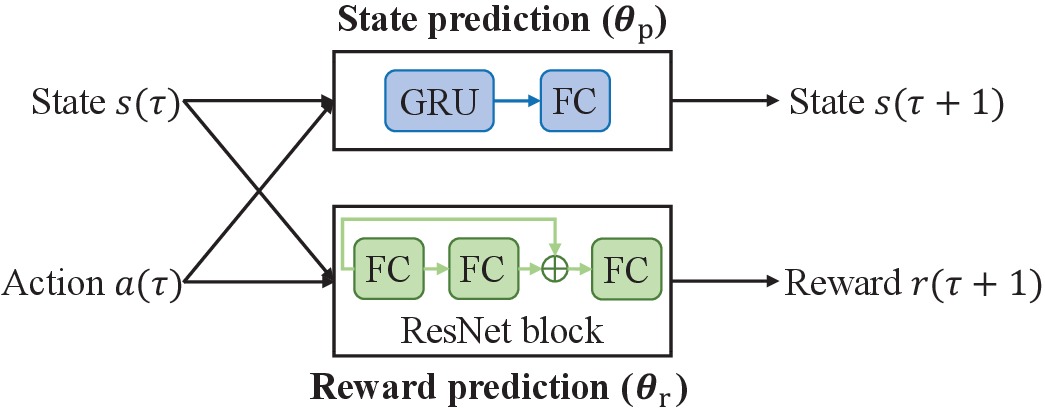}
	\caption{Digital twin module. }
	\label{DTModel}
\end{figure}

The digital twin module provides a pre-validated platform, where the DeepIOS agent learns the efficient IOS parameter configuration by interacting with the virtual environment. By doing so, the overhead resulted from trial-and-error and computation can be significantly reduced, when DeepIOS is deployed in practical systems. Specifically, with the available dataset, the digital twin module is fitted to simulate the practical IOS assisted MU-MIMO environment through supervised learning \cite{caruana2006empirical}. The essence of supervised learning is to learn a mapping from training data to labels. As illustrated in Fig. \ref{DTModel}, a DNN architecture is adopted to construct the digital twin module, which contains: (i) one module for the next state $s({\tau}+1)$ prediction, and (ii) one module for the reward $r({\tau}+1)$ prediction. Furthermore, the input data delivered to the DNN model is the state $s({\tau})$ and the action $a({\tau})$. In other words, both $s({\tau})$ and $a({\tau})$ are the training data, whereas both $s({\tau}+1)$ and $r({\tau}+1)$ are the corresponding labels. Thus, the formulated digital twin module can be described as
\begin{subequations}
\begin{align}\label{stateprediction}
&s({\tau}+1) \sim {\mathcal{P}}_{\boldsymbol{\theta}_{\text{p}}}{\big(}s({\tau}+1)|s({\tau}),a({\tau}){\big)},\\\label{rewardprediction}
&r({\tau}+1) \sim {\mathcal{R}}_{\boldsymbol{\theta}_{\text{r}}}{\big(}s({\tau}),a({\tau}){\big)},
\end{align}
\end{subequations}
where ${\boldsymbol{\theta}_{\text{p}}}$ and ${\boldsymbol{\theta}_{\text{r}}}$ are the parameter vectors of the next state and reward prediction modules, respectively.

Recall that the state $s({\tau})$ is composed of the estimated channel matrix and IOS coefficient matrices, $a({\tau})$ is the amplitude and phase-shift of IOS, and $r({\tau})$ is all UE's sum data rate. The support of the data acquisition module makes it easy to obtain the above data information, as in \eqref{trainingdata}. To distinguish it from DeepIOS's DNN model, the DNN model employed to build the digital twin module is referred to as \textit{supervised DNN}.
If the supervised DNN model can accurately simulate the real MU-MIMO environment, the DeepIOS agent can get massive experience samples for training through ``answering data with data''. In particular, at the beginning (i.e., ${\tau}=0$), the agent decides an action $a(0)$ based on the initialized state $s(0)$; after that, it imports the pair ${\big(}s(0),a(0){\big)}$ into the supervised DNN model, which in turn predicts the new state $s(1)$ and the reward $r(1)$. Subsequently, the agent repeats the above process.

To improve the supervised DNN model's accuracy, one GRU layer followed by one FC layer implement the next state prediction module ${\mathcal{P}}_{\boldsymbol{\theta}_{\text{p}}}$, and one ResNet block \cite{he2016deep} containing three FC layers fits the reward prediction module ${\mathcal{R}}_{\boldsymbol{\theta}_{\text{r}}}$. The main reasons are below. Since the estimated channel matrix is embedded in $s({\tau})$, the state prediction problem in \eqref{stateprediction} is essentially a temporal prediction problem caused by the wireless channel's fast time-varying. Thus, GRU, which has the ability to learn and infer complex temporal correlations, is the best choice for ${\mathcal{P}}_{\boldsymbol{\theta}_{\text{p}}}$. On the contrary, there is just a simple mapping relationship from both $s({\tau})$ and $a({\tau})$ to $r({\tau}+1)$; and thus one ResNet block\footnote{Our previous work \cite{ye2020deep} has exemplified ResNet's efficiency to solve the simple prediction problem in wireless communication systems.} is sufficient to fulfill ${\mathcal{R}}_{\boldsymbol{\theta}_{\text{r}}}$. It should be pointed out that GRU can also implement ${\mathcal{R}}_{\boldsymbol{\theta}_{\text{r}}}$, but its complexity is higher than the ResNet block and it does not yield additional accuracy gain.

To train the supervised DNN model, i.e., update its parameter vectors ${\boldsymbol{\theta}_{\text{p}}}$ and ${\boldsymbol{\theta}_{\text{r}}}$, a mini-batch $\mathcal{B}_{\text{D}}$ composed of $N_{\text{D}}$ latest training data is adopted to calculate the loss function:
\begin{subequations}
\begin{align}\nonumber
&Loss(\boldsymbol{\theta}_{\text{p}})=\frac{1}{N_{\text{D}}}\sum\limits_{c(i){\in}{\mathcal{B}_{\text{D}}}}{\big{\Vert}} {\mathcal{P}}_{\boldsymbol{\theta}_{\text{p}}}{\big(}s(i),a(i){\big)} - s(i+1) {\big{\Vert}}_2^2,\\\nonumber
&Loss(\boldsymbol{\theta}_{\text{r}})=\frac{1}{N_{\text{D}}}\sum\limits_{c(i){\in}{\mathcal{B}_{\text{D}}}}{\big{\Vert}} {\mathcal{P}}_{\boldsymbol{\theta}_{\text{r}}}{\big(}s(i),a(i){\big)} - r(i+1) {\big{\Vert}}_2^2.
\end{align}
\end{subequations}
Furthermore, $\boldsymbol{\theta}_{\text{p}}$ and $\boldsymbol{\theta}_{\text{r}}$ are updated via SGD \cite{sutton2018reinforcement} as follows:
\begin{equation}
    \boldsymbol{\theta}_{\text{p}}{\leftarrow}\boldsymbol{\theta}_{\text{p}}-{{\alpha}_{\text{p}}}{\nabla}Loss(\boldsymbol{\theta}_{\text{p}}),\,\,\,\boldsymbol{\theta}_{\text{r}}{\leftarrow}\boldsymbol{\theta}_{\text{r}}-{{\alpha}_{\text{r}}}{\nabla}Loss(\boldsymbol{\theta}_{\text{r}}),\label{updateparameter2}
\end{equation}
where ${\alpha}_{\text{p}}$ and ${\alpha}_{\text{r}}$ are the learning rates of $\boldsymbol{\theta}_{\text{p}}$ and $\boldsymbol{\theta}_{\text{r}}$, respectively. To build the digital twin module, $N_{\text{D}}$ is set to $D$ during the initialization stage; whereas during the subsequent calibration stage, $N_{\text{D}}$ is smaller than $D$. To prevent over-fitting, the digital twin module is calibrated every $T_1$ time slots.

\subsection{DeepIOS Module}
There is a different DeepIOS module in both the physical space and the digital space. To distinguish them in the following illustration, the DeepIOS module with the parameter vectors ${\boldsymbol{\theta}_{\text{s},\text{d}}}$, ${\boldsymbol{\theta}_{1, \text{d}}}$, and ${\boldsymbol{\theta}_{2, \text{d}}}$ in the digital space is referred to as the \textit{digital DeepIOS}, whilst the DeepIOS module with the parameter vectors ${\boldsymbol{\theta}_{\text{s},\text{p}}}$, ${\boldsymbol{\theta}_{1, \text{p}}}$, and ${\boldsymbol{\theta}_{2, \text{p}}}$ in the physical space is called the \textit{physical DeepIOS}.

To generate experience samples, the digital DeepIOS module frequently interacts with the virtual MU-MIMO environment, i.e., the digital twin module. Specifically, based on the state $s({\tau})$, the digital DeepIOS module decides an appropriate action $a({\tau})$, and then inputs the state $s({\tau})$ and action $a({\tau})$ into the digital twin module. After obtaining the next state $s({\tau}+1)$ and reward $r({\tau}+1)$ delivered by the digital twin module, the digital DeepIOS module stores the experience $e({\tau})={\big(}s({\tau}),a({\tau}),r({\tau}+1),s({\tau}+1){\big)}$ into the experience buffer $\mathcal{E}$. For subsequent training, it updates ${\boldsymbol{\theta}_{\text{s},\text{d}}}$, ${\boldsymbol{\theta}_{1, \text{d}}}$, and ${\boldsymbol{\theta}_{2, \text{d}}}$ by \eqref{loss1} and \eqref{parameterupdate}; and every $T_0$ trainings, it updates the target DNN's parameter vectors ${\boldsymbol{\theta}_{\text{s},\text{d}}^-}$, ${\boldsymbol{\theta}_{1, \text{d}}^-}$, and ${\boldsymbol{\theta}_{2, \text{d}}^-}$ to ${\boldsymbol{\theta}_{\text{s},\text{d}}}$, ${\boldsymbol{\theta}_{1, \text{d}}}$, and ${\boldsymbol{\theta}_{2, \text{d}}}$, respectively.
Thanks to the abundant computing resources in the digital space, the interaction between the digital DeepIOS module and the digital twin module can be completed extremely quickly, making it easy to generate a large number of experiences in a short time. Furthermore, the DNN training will also be effortless. To enable the IOS controller to effectively configure IOS coefficient in the physical space, the digital DeepIOS module delivers the trained evaluation DNN's parameter vectors to update the physical DeepIOS module in real time, i.e.,
\begin{equation}
    {\boldsymbol{\theta}_{\text{s},\text{p}}}{\leftarrow}{\boldsymbol{\theta}_{\text{s},\text{d}}},\,\,\,{\boldsymbol{\theta}_{1, \text{p}}}{\leftarrow}{\boldsymbol{\theta}_{\text{1},\text{d}}},\,\,\,{\boldsymbol{\theta}_{2, \text{p}}}{\leftarrow}{\boldsymbol{\theta}_{2, \text{d}}}.\label{updateparameter3}
\end{equation}

Unlike the digital DeepIOS module with both evaluation DNN and target DNN, the physical DeepIOS module is only equipped with an evaluation DNN for configuring the IOS parameter. Due to the lack of target DNN, the physical DeepIOS module cannot train the evaluation DNN. In other words, the physical DeepIOS module performs the strategy in an offline manner: it only makes decisions based on the state via a trained evaluation DNN. By dong so, a large amount of computational resource and time, required for DNN training \cite{qian2020noma}, can be saved to ensure real-time decision-making in practical wireless systems (as the evaluation DNN maps from the state to the action in microseconds). Furthermore, the interaction between the physical DeepIOS module and the practical MU-MIMO system is necessary, in that this process will produce fresh data $c(t)$ to calibrate the digital twin module.

\subsection{Operation of DeepIOS with Digital Twins}
As shown in Fig. \ref{DTDeepIOS}, the DeepIOS framework with digital twins contains three closed loops: the digital twin module and the digital DeepIOS module form the first closed-loop; the practical MU-MIMO system and the physical DeepIOS module form the second closed-loop; and the digital space, the physical space, and the data acquisition module form the third closed-loop. These three closed loops promote each other to enhance the whole system's performance. Specifically, in the first closed-loop, the digital DeepIOS module frequently interacts with the digital twin module to generate a large number of experiences for the evaluation DNN training; in the second closed loop, the physical DeepIOS module controls IOS to improve the MU-MIMO system's performance, without prior sub-channels' CSI and UEs' mobility; in the third closed-loop, the digital DeepIOS module delivers the evaluation DNN's parameter vector to the physical DeepIOS module for policy improvement, and meanwhile the data acquisition module collects real data from the practical environment to calibrate the digital twin module. Such a framework has three main advantages that are very important for practical systems.
\begin{itemize}
    \item In the physical space, the IOS and the MU-MIMO system are unaware of each other except for minimal information exchange, thus speeding up the rollout of IOS.
    \item The physical DeepIOS module only makes decisions without online training that consumes time, such that the decision-making's real-time feature can be satisfied.
    \item The digital DeepIOS module's online learning and parameter vector delivery enable the physical DeepIOS module to fit well in the environmental dynamic change.
\end{itemize}

Algorithm \ref{DeepIOS1} provides the pseudocode of DeepIOS with digital twins, and the scheme operations include: (i) initialize all parameters (lines 1$\sim$3); (ii) collect real-time data via \eqref{trainingdata} and store it into the data requisition module (line 6); (iii) construct and calibrate the digital twin module (lines 8$\sim$14); (iv) train the digital DeepIOS module to interact with the digital twin module (lines 16$\sim$29); and (v) update the physical DeepIOS module and enable it to interact with the MU-MIMO system for performance improvements and new data generation (lines 31$\sim$35).

\begin{algorithm}[t]\caption{Digital Twin Enhanced DeepIOS Scheme}\label{DeepIOS1}
	\begin{algorithmic}[1]
		\State Initialize $s(0)$, $\boldsymbol{\Phi}_{\text{r}}(0)$, $\boldsymbol{\Phi}_{\text{t}}(0)$, $ \epsilon _{\text{d}}$, $ \epsilon _{\text{p}}$, $ \gamma $, $\alpha$, $\mathcal E$, $N_{\text{E}}$, $ T_0$, $\Gamma$.
              \State Initialize ${\boldsymbol{\theta}_{\text{s},\text{d}}}$, ${\boldsymbol{\theta}_{1, \text{d}}}$, ${\boldsymbol{\theta}_{2, \text{d}}}$, ${\boldsymbol{\theta}_{\text{s},\text{d}}^-}$, ${\boldsymbol{\theta}_{1, \text{d}}^-}$, ${\boldsymbol{\theta}_{2, \text{d}}^-}$, ${\boldsymbol{\theta}_{\text{s},\text{p}}}$, ${\boldsymbol{\theta}_{1, \text{p}}}$, ${\boldsymbol{\theta}_{2, \text{p}}}$.
		\State Initialize $R_{\text{th}}$, ${\alpha _{\text{p}}}$, ${\alpha _{\text{r}}}$, $\mathcal D$, $N_{\text{D}}$,  ${\boldsymbol{\theta}_{\text{p}}}$, ${\boldsymbol{\theta}_{\text{r}}}$, $T_1$.
		\For{$ t=0,1,2, \cdots $}
          \State \textbf{Data Requisition Module:}
              \State Collect real-time data via \eqref{trainingdata}.
          \State \textbf{Digital Twin Module:}
              \If{$ t=0$}
              \State Acquire all training data from $\mathcal D$;
              \State Train ${\boldsymbol{\theta}_{\text{p}}}$ and ${\boldsymbol{\theta}_{\text{r}}}$ via \eqref{updateparameter2}.
              \ElsIf{$\text{Remainder}(t/{T_1})=0$}
              \State Acquire $N_{\text{D}}$ latest training data from $\mathcal D$;
              \State Calibrate ${\boldsymbol{\theta}_{\text{p}}}$ and ${\boldsymbol{\theta}_{\text{r}}}$ via \eqref{updateparameter2}.
              \EndIf
          \State \textbf{Digital DeepIOS Module:}
            \For{${\tau}=0,1,\cdots,\Gamma$}
		\State Input $s(\tau)$ into the evaluation DNN;
            \State Obtain the Q-values for different actions $a{\in}{\mathcal{A}}$;
		\State Choose $a(\tau)={\big(}{a_1}(\tau),{a_2}(\tau){\big)}$ via \eqref{actionselection} with ${\epsilon}_{\text{d}}$;
		\State Input $s(\tau)$ and $a(\tau)$ into the digital twin module;
            \State Obtain $s({\tau}+1)$ via \eqref{stateprediction} and $r({\tau}+1)$ via \eqref{rewardprediction};
          \State  Store ${\big(}s(\tau),a(\tau),r({\tau}+1),s({\tau}+1){\big)}$ into $\mathcal{E}$;
          \State Sample $N_{\text{E}}$ experiences from $\mathcal{E}$ to form $\mathcal{B}_{\text{E}}$;
          \State Calculate $Loss({\boldsymbol{\theta}_{\text{s},\text{d}}}$, ${\boldsymbol{\theta}_{1, \text{d}}}$, ${\boldsymbol{\theta}_{2, \text{d}}})$ via \eqref{loss1};
		\State Train ${\boldsymbol{\theta}_{\text{s},\text{d}}}$, ${\boldsymbol{\theta}_{1, \text{d}}}$, and ${\boldsymbol{\theta}_{2, \text{d}}}$ via \eqref{parameterupdate}.
            \If{$\text{Remainder}({\tau}/{T_0})=0$}
              \State ${\boldsymbol{\theta}_{\text{s},\text{d}}^-}{\leftarrow}{\boldsymbol{\theta}_{\text{s},\text{d}}}$, ${\boldsymbol{\theta}_{1, \text{d}}^-}{\leftarrow}{\boldsymbol{\theta}_{1, \text{d}}}$, ${\boldsymbol{\theta}_{2, \text{d}}^-}{\leftarrow}{\boldsymbol{\theta}_{2, \text{d}}}$.
              \EndIf
        \EndFor
         \State \textbf{Physical DeepIOS Module:}
		\State Update ${\boldsymbol{\theta}_{\text{s},\text{p}}}$, ${\boldsymbol{\theta}_{1, \text{p}}}$, and ${\boldsymbol{\theta}_{2, \text{p}}}$ via \eqref{updateparameter3};
            \State Obtain $\widetilde{\textbf{H}}(t)$ from BS and form $s(t)$;
            \State Input $s(t)$ into the evaluation DNN to obtain Q-values;
            \State Choose $a(t)={\big(}{a_1}(t),{a_2}(t){\big)}$ via \eqref{actionselection} with ${\epsilon}_{\text{p}}$;
            \State Take $a(t)$ to interact the environment.
		\EndFor
	\end{algorithmic}
\end{algorithm}

\subsection{Computational Complexity Analysis}
As shown in Fig. \ref{DeepIOS2}, the DNN model of DeepIOS consists of three input layers, three GRU layers, five FC layers, one concatenation layer, and two output layers. The DNN model of the digital twin model consists of the next state prediction module and the reward prediction module, where the next state prediction module contains one input layer, one GRU layer, one FC layer, and one output layer, while the reward prediction module contains one input layer, three FC layers, and one output layer. According to \cite{cho2014learning}, the computational complexity of GRU layer $i$ for forward-propagation and back-propagation is $3h_{i-1} h_i^2$, wherein $h_{i-1}$ and $h_i$ are the number of neurons in GRU layer $i$ and its previous layer, respectively. According to \cite{DeepLearning}, the computational complexity of FC layer $i$ for forward-propagation is $g_{i-1} g_i$, and its computational complexity for back-propagation is $g_i g_{i+1}$, wherein $g_{i-1}$, $g_i$, and $g_{i+1}$ are the number of neurons in GRU layer $i$, its previous layer, and its next layer, respectively. Similarly, the computational complexity of concatenation layer $i$ for forward-propagation and back-propagation is given by $f_{i-1} f_i$ and $f_i f_{i+1}$, respectively, wherein $f_{i-1}$, $f_i$, and $f_{i+1}$ denote the number of neurons in concatenation layer $i$, its previous layer, and its next layer, respectively. In this paper, the number of neurons in all FC layers is the same, while only one GRU layer, one FC layer, and one concatenation layer are required. Thus, we omit the subscript $i$ in the following analysis.

We analyze the computational complexity of DeepIOS in the physical space and the digital space, respectively. In the physical space, the physical DeepIOS module only makes decisions without training the DNN, in the sense that its computational complexity is dominated by forward-propagation, i.e., $\mathcal{O}(3h^2(4M^2+2NK)+3hg+3gf+2fg+g(L_1+L_2 ))=\mathcal{O}(6h^2(2M^2+NK)+g(3h+5f+L_1+L_2 )$. In the digital space, (i) the digital DeepIOS module needs to make decisions and train the DNN, and thus its computational complexity is dominated by both forward-propagation and back-propagation, i.e., $\mathcal{O}(3h^2 (4M^2+2NK)+3hg+3gf+2fg+g(L_1+L_2 ))+\mathcal{O}(h(4M^2+2NK)+3h^2 (4M^2+2NK)+3gf+2fg+g(L_1+L_2 ))=\mathcal{O}(2h(6h+1)(2M^2+NK)+2g(5f+L_1+L_2+h))$; and (ii) similarly, the digital twin module needs to predict the next state and the reward, as well as train the DNN, and thus its computational complexity is calculated by $\mathcal{O}(3h^2 (4M^2+2NK)+hg+g(4M^2+2NK))+\mathcal{O}(h(4M^2+2NK)+3h^2 (4M^2+2NK)+g(4M^2+2NK))+\mathcal{O}(g(4M^2+2NK)+g^2+g(4M^2+2NK+g)+g)+\mathcal{O}(g(4M^2+2NK)+g^2+g(4M^2+2NK+g)+g)=\mathcal{O}(2(2M^2+NK)(6h^2+4g+h)+g(h+2g+1))$. Therefore, the total computational complexity of DeepIOS in the digital space is $\mathcal{O}(2(2M^2+NK)(6h^2+h)+2g(5f+L_1+L_2+h))+\mathcal{O}(2(2M^2+NK)(6h^2+4g+h)+g(h+2g+1))=\mathcal{O}(4(2M^2+NK)(6h^2+2g+h)+g(3h+2g+1+10f+2L_1+2L_2))$. Overall, in both physical and digital spaces, the total computational complexity of DeepIOS is $\mathcal{O}(6h^2 (2M^2+NK)+g(3h+5f+L_1+L_2 )+\mathcal{O}(4(2M^2+NK)(6h^2+2g+h)+(3h+2g+1+10f+2L_1+2L_2)g)=\mathcal{O}(2(2M^2+NK)(15h^2+4g+2h)+(6h+2g+1+15f+3L_1+3L_2)g)$.

As analyzed, the computational complexity of DeepIOS in the digital space is significantly higher than that in the physical space. Actually, the essence of DeepIOS with digital twins is to transfer the computational cost from the physical space to the digital space, thereby ensuring real-time decision-making of the physical DeepIOS module. In this regard, the operational cost of using digital twins is large amounts of computational resource required to train the digital twin and digital DeepIOS modules in the digital space. For realistic implementation, the digital space, including the digital twin module and the digital DeepIOS module, can be deployed on the edge server connected to both the BS and the IOS controller via wire \cite{fan2021digital}.

\section{Performance Evaluation}\label{4}
This section evaluates the performances of various schemes based on Python 3.6 simulation platform. The keras library \cite{keras} builds various DNN models, and the py-itpp library \cite{pyitpp} implements communications and signal processing.

\subsection{Parameter Settings}
\subsubsection{System Setup}
We consider a three-dimensional coordinate system, where the BS with a uniform linear array is located at $\textbf{p}_{\text{B}}=[0,0,10]$m, the IOS with a uniform linear array is located at $\textbf{p}_{\text{O}}=[-2,5,5]$m, and five single-antenna UEs including three reflected UEs and two refracted UEs. Each time slots, each UE's location is randomly initialized in the area $[-10, 10)$m $\times$ $[-10, 10)$m with the height being $1.5$m. To model the real-world UE mobility, we adopt the Gauss-Markov mobility model \cite{tabassum2019fundamentals} to generate the velocity and movement direction of each UE, where the memory level parameter is set to $0.5$, the asymptotic mean and of the velocity are $1$m/sec and $0.5$m/sec, respectively, the asymptotic mean and standard deviation of the movement direction are $30^{\circ}$ and $10^{\circ}$, respectively. The BS's antenna number $N$ is 5, and the IOS consists of $M=32$ elements. The noise variance ${\sigma}_{\text{p}}^2$ at BS is 0.1dBm/Hz, and ${\sigma}_k^2$ at UE is 0.5dBm/Hz for $\forall\,k{\in}{\mathcal{K}}$ \cite{wang2022intelligent}. The optional phase-shift increments and amplitudes of IOS will be detailed in each subsection.
The LoS channel between BS and IOS is $\textbf{H}_{\text{BO}}^{\text{LOS}}=\textbf{u}_{\text{O}}\textbf{u}_{\text{B}}^H$, wherein the steering vectors are ${\textbf{u}_{\text{O}}(\psi_{\text{O}})}={\big{[}}1,{e^{j{\pi}{\psi_{\text{O}}}}},\cdots,$ ${e^{j(M-1){\pi}{\psi_{\text{O}}}}}{\big{]}}^{T}$ and ${\textbf{u}_{\text{B}}(\psi_{\text{B}})}={\big{[}}1,$ ${e^{j{\pi}{\psi_{\text{B}}}}},\cdots,{e^{j(N-1){\pi}{\psi_{\text{B}}}}}{\big{]}}^{T}$.
According to \cite{wang2022intelligent}, the directional cosines $\psi_{\text{O}}$ and $\psi_{\text{B}}$ are determined by the relative position between BS and IOS. Suppose that the IOS and the BS are placed along y-axis and x-axis, respectively, thus $\psi_{\text{O}}=$ $[0,1,0]({\textbf{p}_{\text{B}}}-{\textbf{p}_{\text{O}}})/{\Vert {\textbf{p}_{\text{B}}}-{\textbf{p}_{\text{O}}} \Vert}_2$ and $\psi_{\text{B}}=[1,0,0]({\textbf{p}_{\text{B}}}-{\textbf{p}_{\text{O}}})/$ ${\Vert {\textbf{p}_{\text{B}}}-{\textbf{p}_{\text{O}}} \Vert}_2$.
The NLoS components follow the Gaussian distribution, i.e., $\textbf{H}_{\text{BO}}^{\text{NLOS}}\,{\in}\,{\mathcal{CN}}(0,1)$. The channel matrix $\textbf{H}_{\text{BU}}$ and $\textbf{H}_{\text{OU}}$ are generated in the similar way, and the Rician factor $\lambda$ is 10 unless stated otherwise.

\subsubsection{Algorithm Setup}
Fig. \ref{DeepIOS2} and Fig. \ref{DTModel} present the DeepIOS and digital twin modules' DNN models, respectively, where the number of neurons in all layers is 64, the activation function in the DNN output layer of digital twins and other DNN models are Linear and ReLU \cite{sutton2018reinforcement}, respectively. TABLE \ref{tab:table2} summarizes the detailed parameters of DeepIOS with digital twins. The initialization data in the data acquisition module (i.e., the historical data for training the digital twin module) is obtained under the case where $\lambda=9$ and UEs are randomly distributed. When the algorithm begins to execute, the above parameters and NLoS components are reconfigured. In the simulation, the following algorithms are considered:
\begin{itemize}
    \item Random scheme: the parameters of IOS, i.e., phase-shift and amplitude, are chosen in random in each time slot.
    \item MAB scheme: the parameters of IOS are selected through multi-arm bandit (MAB) \cite{sutton2018reinforcement}; specifically, the Thompson Sampling algorithm is adopted in this scheme \cite{wang2022intelligent}.
    \item DeepIOS scheme: this is the proposed DRL based IOS control scheme in this paper.
    \item DeepIOS with digital twins scheme: this is our proposed IOS control scheme based on DRL with digital twins.
\end{itemize}

The optimization based method is not suitable as the benchmark, since (i) sub-channels' CSI is not available and (ii) the wireless channel and all UEs' locations vary in each time slot.

\begin{table*}[]
    \centering
    \caption{Algorithm parameters.}
    \begin{tabular}{|c|c|c|c|c|c|}\hline
     Parameter & Value & Parameter & Value & Parameter & Value\\\hline
     Learning rate ($\alpha$)  & 0.001 & Mini-batch size ($N_{\text{E}}$) & 8 & Calibration frequency ($T_1$) & 10\\\hline
     Learning rate (${\alpha _{\text{p}}}$) & 0.001 & Mini-batch size ($N_{\text{D}}$) & 24 & Exploring probability ($ \epsilon _{\text{d}}$) & $\max\{0.001,0.99^{\tau}\}$\\\hline
     Learning rate (${\alpha _{\text{r}}}$) & 0.001 & Experience buffer size ($E$) & 10000 & Exploring probability ($ \epsilon _{\text{p}}$) & $\max\{0.001,0.99^{t}\}$\\\hline
     Discount factor ($ \gamma $) & 0.95 & Digital twin dataset size ($D$) & 1000 & Penalty factor ($\omega$) & 20 unless stated otherwise\\\hline
     Reward threshold ($R_{\text{th}}$) & 10 & Update frequency ($ T_0$) & 20 & Number of interactions ($\Gamma$) & 1000 unless stated otherwise\\\hline
    \end{tabular}
    \label{tab:table2}
\end{table*}

\subsubsection{Metric Setup}
The performance metrics include data rate, convergence time, and simulation run-time. To be specific, the system's sum data rate at time slot $t$ is a short-term average\footnote{A key advantage of adopting the ``short-term average data rate'' as the performance metric is that it can avoid the impact of low data rate generated in the process of trial-and-error interaction with the environment (i.e., the initial stage of simulation) on the subsequent data rate \cite{ye2021multi}.}, which is given by $\sum\nolimits_{l=t-{N_{\text{r}}}+1}^t{\big(}\sum\nolimits_{k=1}^{K}{R_k(l)}{\big)}/{N_{\text{r}}}$, wherein $N_{\text{r}}$ is 2000 and $R_k(l)$ is in \eqref{datarate}; the convergence time is defined as the time slots required for the algorithm to converge; and the simulation run-time is defined as the actual average running time of the simulation program in each simulated time slot. To ensure the experimental accuracy, all simulations run 20 times to obtain the average result.

\subsection{Impact of Different Rician Factors}\label{Exp1}
\begin{figure*}[t]
	\centering
    \subfloat[\label{0a} $\lambda=5$.]{
    \begin{minipage}[b]{0.326\textwidth}
	\includegraphics[scale=0.396]{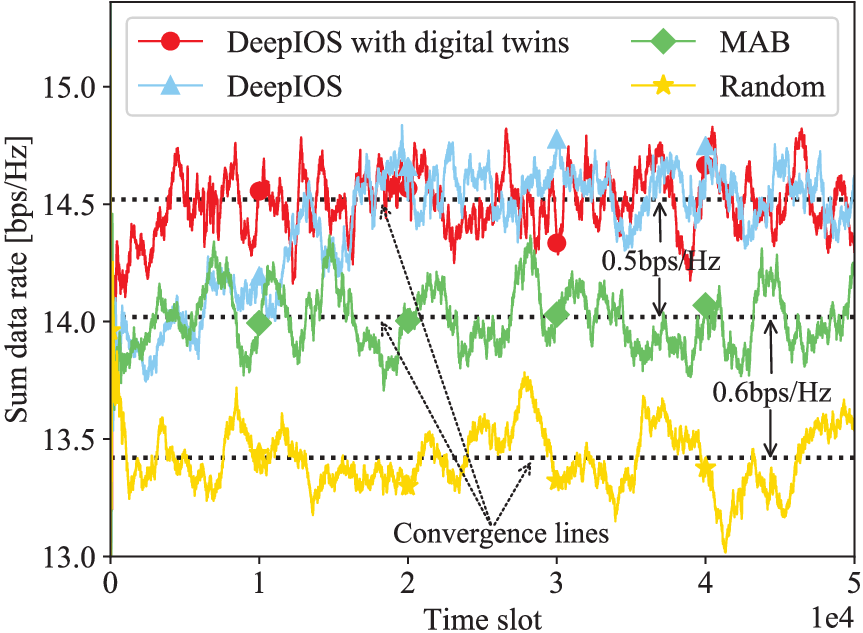}
    \end{minipage}}
    \subfloat[\label{0b} $\lambda=10$.]{
    \begin{minipage}[b]{0.326\textwidth}
	\includegraphics[scale=0.396]{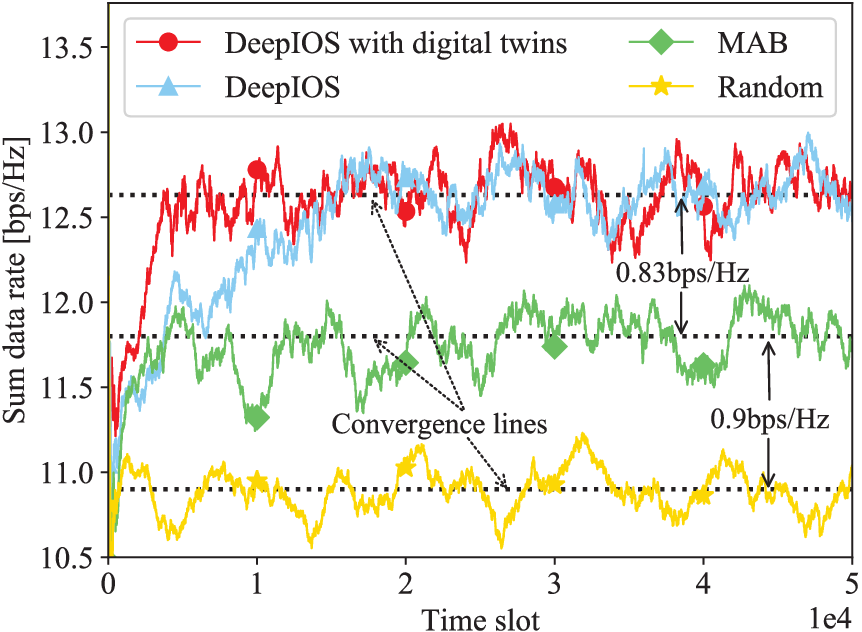}
    \end{minipage}}
    \subfloat[\label{0c} $\lambda=15$.]{
    \begin{minipage}[b]{0.326\textwidth}
	\includegraphics[scale=0.396]{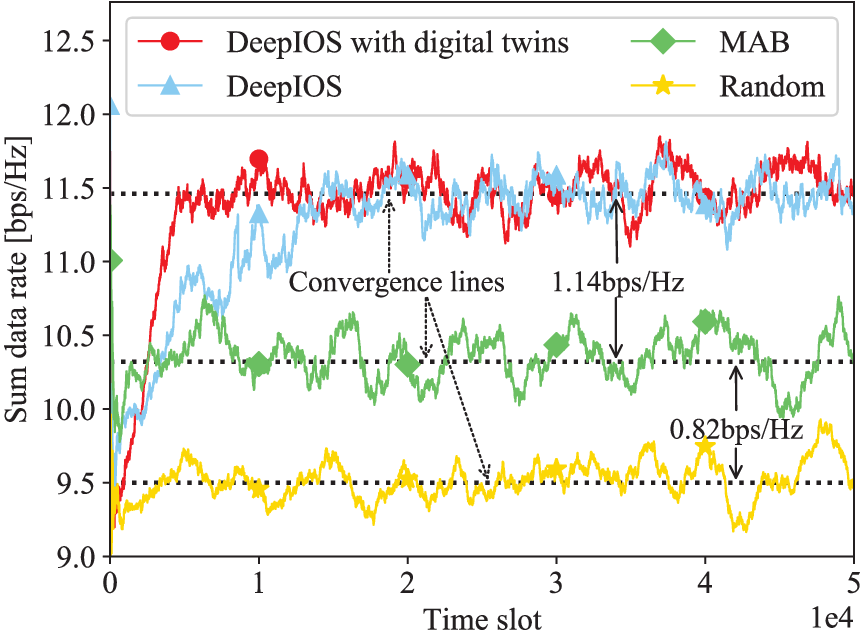}
    \end{minipage}}
	\caption{Sum data rates attained by various schemes under ES-IOS.}
	\label{Fig0}
\end{figure*}

\begin{figure*}[t]
	\centering
    \subfloat[\label{1a} $\lambda=5$.]{
    \begin{minipage}[b]{0.326\textwidth}
	\includegraphics[scale=0.396]{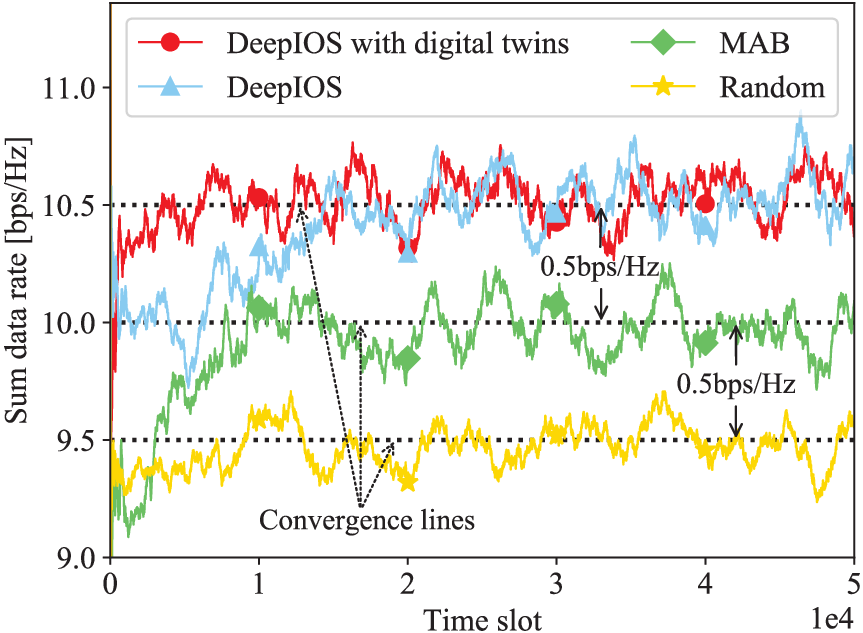}
    \end{minipage}}
    \subfloat[\label{1b} $\lambda=10$.]{
    \begin{minipage}[b]{0.326\textwidth}
	\includegraphics[scale=0.396]{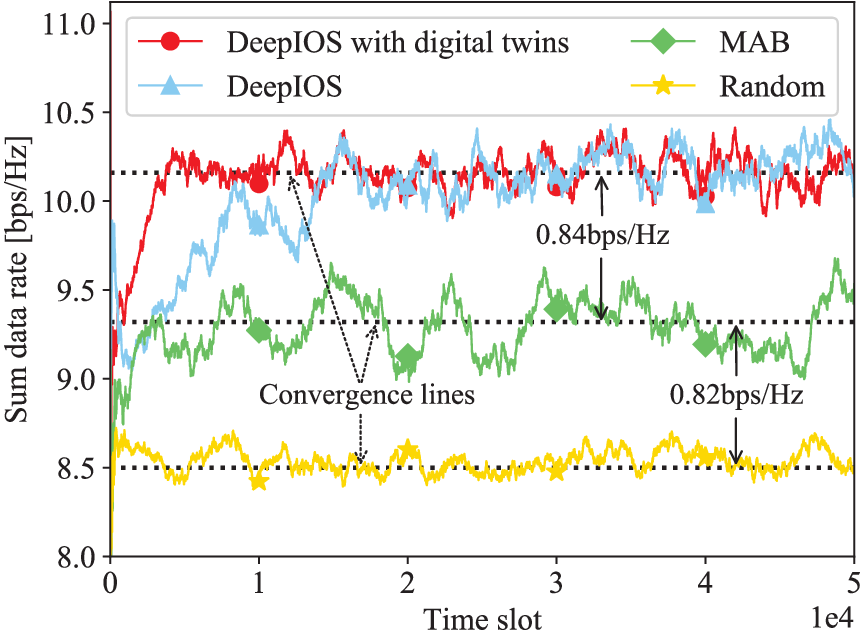}
    \end{minipage}}
    \subfloat[\label{1c} $\lambda=15$.]{
    \begin{minipage}[b]{0.326\textwidth}
	\includegraphics[scale=0.396]{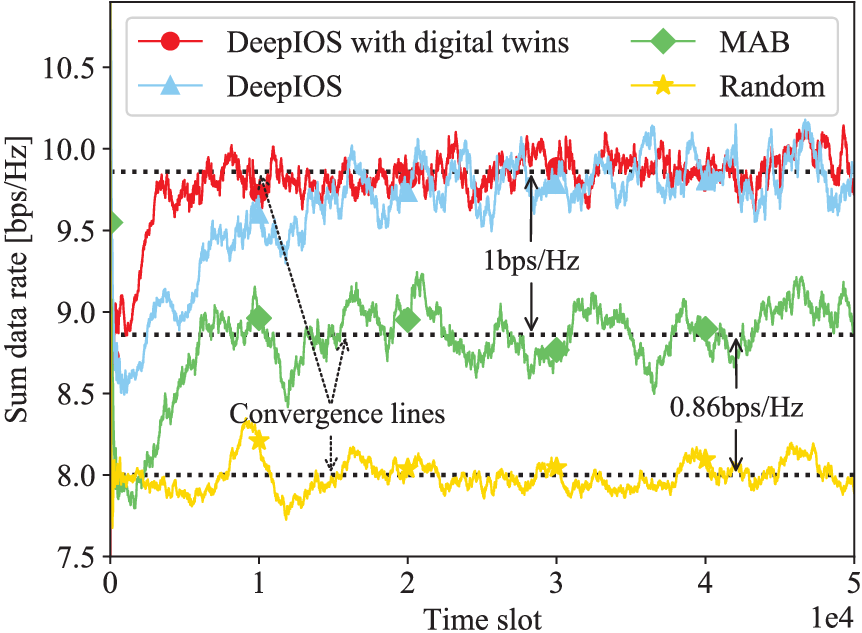}
    \end{minipage}}
	\caption{Sum data rates attained by various schemes under MS-IOS.}
	\label{Fig1}
\end{figure*}

This subsection evaluates the impact of different Rician factors on various schemes' performance, in which $\lambda$ varies from 5 to 15 with a step size of 5. Fig. \ref{Fig0} and Fig. \ref{Fig1} provide the sum data rates attained by different schemes under ES-IOS and MS-IOS, respectively. To clear the convergence process, the convergence lines (i.e., the lines along which various schemes tend to be stable in the long-term average) are marked. It can be seen that due to wireless channels' fast time-varying and UEs' random mobility, all the curves are in great fluctuation.

\textit{1) Case of ES-IOS:} the phase-shift increment set ${\mathcal{A}}_1$ is ${\big{\{}}\textbf{w}(-3/M),$ $\textbf{w}(-1/M),\textbf{w}(0),\textbf{w}(1/M),\textbf{w}(3/M){\big{\}}}$, and the ratio of ${\beta}_{\text{r},m}^2$ to ${\beta}_{\text{t},m}^2$ can be selected from $\{100,10,1,0.1,$ $0.01\}$, i.e., the reflecting amplitude set is ${\mathcal{A}}_2=\{0.995,0.953,$ $0.707,0.302,0.100\}$. As shown in Fig. \ref{Fig0}, all schemes do not rely on sub-channels' CSI, but their utilization of other information is different. As expected, the random scheme achieves the worst performance since its decisions are independent of any information. In contrast, the MAB scheme first obtains all optional actions' reward distribution through trial-and-error; subsequently, it selects the action that obtains the maximum reward to execute at each time slot. Thus, compared with the random scheme, the MAB scheme yields a more than 0.6bps/Hz higher data rate for all simulated $\lambda$. MAB, however, is a stateless algorithm; in other words, it fails to describe the environmental state and establish the connection between its action and the environment. It consequently addresses the issue in simple scenarios, instead of the system where the UEs' mobility is random and the channels' CSI is fast time-varying. Unlike MAB, DeepIOS's underlying technique is DRL with the state setup and thus it can learn the state-action pair's quality according to the reward. Owing to its ingenious definition of the state, DeepIOS configures more appropriate parameters for IOS. Therefore, in comparison to MAB, DeepIOS further improves the system's data rate: the data rate gain is more than 0.5bps/Hz for various $\lambda$.

Fig. \ref{Fig0} also presents two interesting phenomenon. (i) As $\lambda$ increases, all schemes' performance decreases. When $\lambda$ changes from 5 to 15, DeepIOS's data rate reduction is about 3.06bps/Hz, which indicates that its effectiveness is dependent on the external environment. (ii) As $\lambda$ increases, the data rate gap between DeepIOS and other schemes increases. When $\lambda$ varies from 5 to 15, the gain of the data rate gap between DeepIOS and the random scheme is around 96\%, and that between DeepIOS and the MAB scheme is about 128\%. This observation demonstrates that compared with other schemes, DeepIOS is more robust to different scenarios.

\begin{table}[]
    \centering
    \caption{Time slots [$\times{1e4}$] required for DeepIOS and DeepIOS with digital twins to converge under different $\lambda$ in both ES-IOS and MS-IOS cases.}
    \begin{tabular}{|c|c|c|c|}\hline
     \multicolumn{4}{|c|}{ES-IOS}\\\hline
      & $\lambda=5$ & $\lambda=10$ & $\lambda=15$\\\hline
     DeepIOS with digital twins & 0.39 & 0.39 & 0.42\\\hline
     DeepIOS & 1.38 & 1.51 & 1.47\\\hline
     \multicolumn{4}{|c|}{MS-IOS}\\\hline
      & $\lambda=5$ & $\lambda=10$ & $\lambda=15$\\\hline
     DeepIOS with digital twins & 0.37 & 0.38 & 0.37\\\hline
     DeepIOS & 1.45 & 1.49 & 1.50\\\hline
    \end{tabular}
    \label{tab:table3}
\end{table}

TABLE \ref{tab:table3} summarizes the time slots required for DeepIOS and DeepIOS with digital twins to converge under different $\lambda$ in both ES-IOS and MS-IOS cases. As expected, benefiting from pre-interaction in the digital space, DeepIOS with digital twins achieves faster convergence than DeepIOS: for all simulated $\lambda$, compared with DeepIOS, DeepIOS with digital twins consumes at least 71.4\% less time slots to converge. On the other hand, we can find from Fig. 6 that under different $\lambda$, the data rate of DeepIOS is comparable to that of DeepIOS with digital twins, and the two have the same data rate change trend. These results confirm that the digital twin module enables DeepIOS to avoid too many unnecessary trial-and-error interactions, thereby accelerating the convergence speed.

\textit{2) Case of MS-IOS:} the phase-shift increment set ${\mathcal{A}}_1$ is ${\big{\{}}\textbf{w}(-3/M),$ $\textbf{w}(-1/M),\textbf{w}(0),\textbf{w}(1/M),\textbf{w}(3/M){\big{\}}}$, and five types of element groups are considered, each of which initializes ${\beta}_{\text{r},m}$, $\forall\,\,m{\in}\mathcal{M}$ to be 0 or 1 in random. Similar to the ES-IOS case, in the MS-IOS case, (i) DeepIOS attains higher data rates compared with MAB and random schemes for all considered $\lambda$; (ii) DeepIOS shows stronger robustness than MAB and random schemes; and (iii) the digital twin module improves DeepIOS's convergence speed.

Furthermore, as depicted in Fig. \ref{Fig1} various algorithms' data rates under ES-IOS are always better than that under MS-IOS: for all $\lambda$, the data rate gains of DeepIOS with digital twins, DeepIOS, MAB, and random schemes are more than 16.2\%, 16.2\%, 16.5\%, and 18.8\%, respectively. This is because that the MS protocol's amplitude coefficient is limited to binary variables, and it thus cannot flexibly control the reflecting and refracting coefficients. On the contrary, the ES protocol dynamically adjusts the elements' amplitude to enable simultaneous reflection and refraction at one time. As a result, ES-IOS can take full advantage of the degrees-of-freedoms available to each element to enhance the desired signal. On the other hand, this phenomenon can also be explained by the fact that the MS protocol is the ES protocol's special case.

\subsection{Impact of Different Sub-action Sets}\label{Exp2}
This simulation studies the impact of different sub-action sets on DeepIOS's performance when $\lambda=10$. Furthermore, to evaluate the action branch architecture's effectiveness, the DeepIOS scheme without the action branch architecture is employed as the benchmark.

\begin{figure}[t]
	\centering
    \subfloat[\label{2a} Under the ES-IOS case.]{
    \begin{minipage}[b]{0.46\textwidth}
	\includegraphics[scale=0.446]{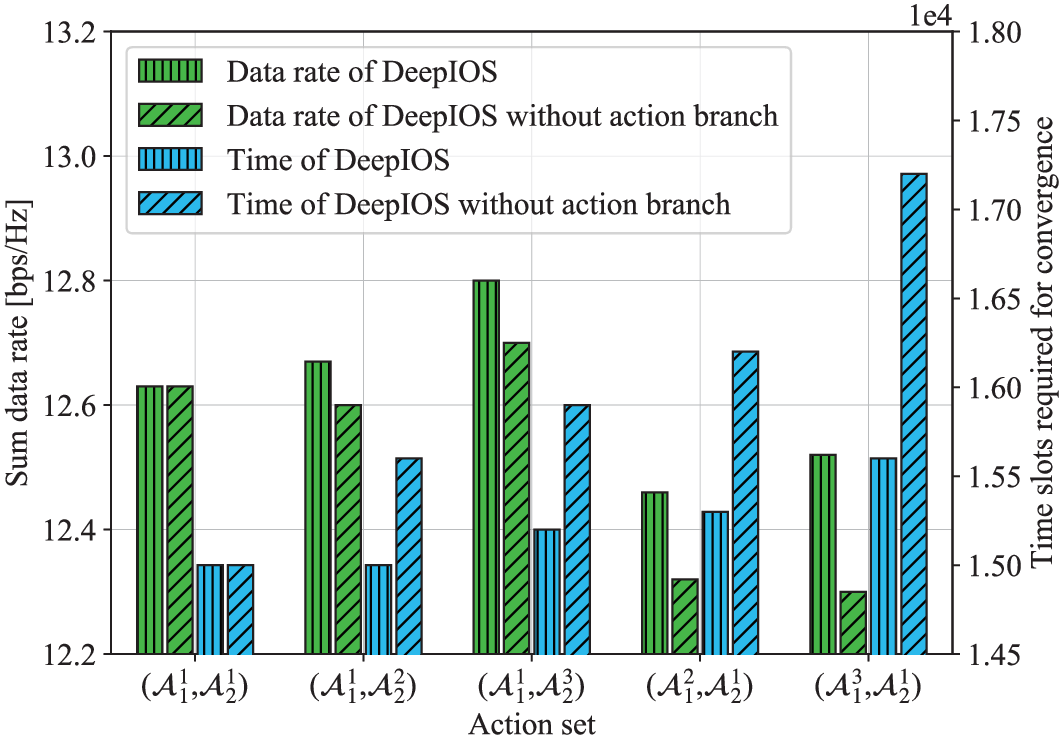}
    \end{minipage}}
   \\
    \subfloat[\label{2b} Under the MS-IOS case.]{
    \begin{minipage}[b]{0.46\textwidth}
	\includegraphics[scale=0.446]{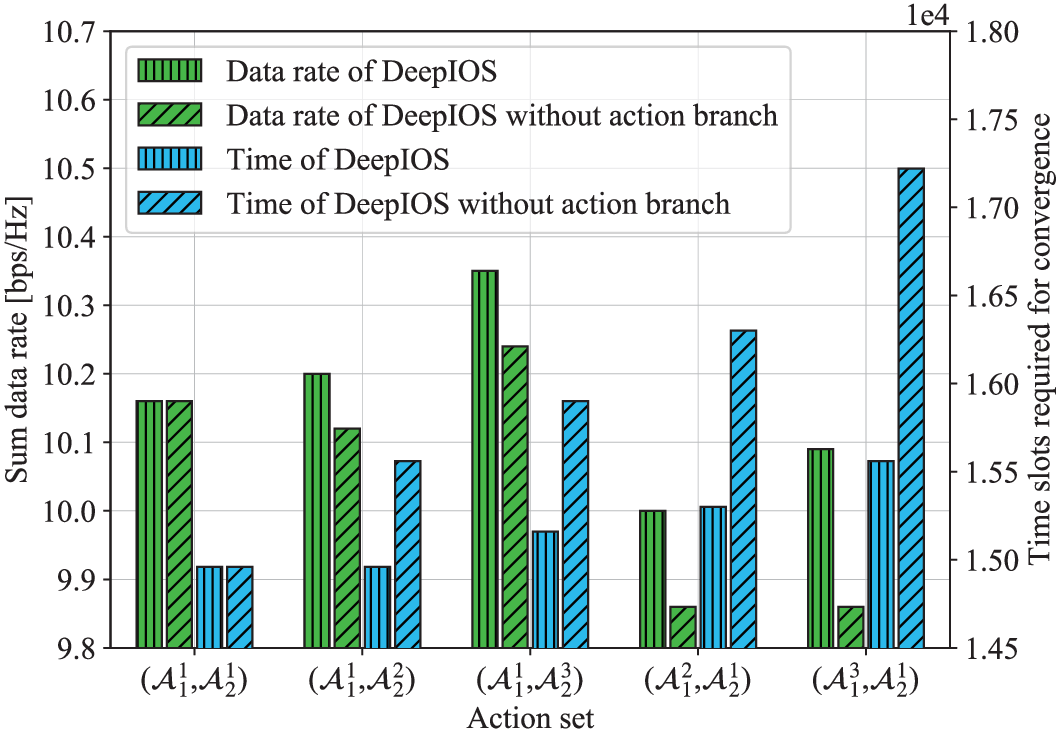}
    \end{minipage}}
	\caption{Sum data rates and convergence time slots of various schemes under both ES-IOS and MS-IOS cases.}
	\label{Fig2}
\end{figure}

\textit{1) Case of ES-IOS:} the following five types of sub-action sets are considered, i.e., $({\mathcal{A}_1^1},{\mathcal{A}_2^1})$, $({\mathcal{A}_1^1},{\mathcal{A}_2^2})$, $({\mathcal{A}_1^1},{\mathcal{A}_2^3})$, $({\mathcal{A}_1^2},{\mathcal{A}_2^1})$, $({\mathcal{A}_1^3},{\mathcal{A}_2^1})$, where
${\mathcal{A}_1^1}={\big{\{}}$\textbf{w}($-3/M$), \textbf{w}$(-1/M),$ \textbf{w}(0)$,$ \textbf{w}$(1/M),$ \textbf{w}$(3/M){\big{\}}}$, ${\mathcal{A}_1^2}={\big{\{}}$\textbf{w}$(-9/M),$ $\cdots,$ \textbf{w}$(-2/M),$ \textbf{w}$(-1/M),$ \textbf{w}(0)$,$ \textbf{w}$(1/M),$ \textbf{w}$(2/M),$ $\cdots,$ \textbf{w}$(9/M){\big{\}}}$, ${\mathcal{A}_1^3}={\big{\{}}$ \textbf{w}$(-15/M),$ $\cdots,$ \textbf{w}$(-2/M),$ \textbf{w}$(-1/M),$ \textbf{w}(0)$,$ \textbf{w}$(1/M),$ \textbf{w}$(2/M),$ $\cdots,$ \textbf{w}$(15/M){\big{\}}}$, ${\mathcal{A}_2^1}=\{0.995,$ $0.953,$ $0.707,$ $0.302,$ $0.100\}$, ${\mathcal{A}_2^2}=\{0.998,$ $0.995,$ $0.990,$ $0.953,$ $0.913,$ $0.707,$ $0.577,$ $0.302,$ $0.218,$ $0.100\}$, and ${\mathcal{A}_2^3}=\{1.000,$ $0.999,$ $0.998,$ $0.995,$ $0.990,$ $0.953,$ $0.913,$ $0.806,$ $0.707,$ $0.577,$ $0.302,$ $0.218,$ $0.100,$ $0.070,$ $0.032\}$. In the above settings, the values of ${\mathcal{A}_1^1}$, ${\mathcal{A}_1^2}$, and ${\mathcal{A}_1^3}$ are similar to \cite{wang2022intelligent}, where the discrete Fourier transform vectors are partially used as the optional phase-shift increments. The values of ${\mathcal{A}_2^1}$, ${\mathcal{A}_2^2}$, and ${\mathcal{A}_2^3}$ are similar to \cite{zhang2022intelligent1}, where the ratio of the reflecting amplitude to the refracting amplitude is given in the form of different quantities, and the details are provided in Section \ref{Exp1}-1.

As can be seen from Fig. \ref{2a}, both DeepIOS schemes achieve the highest data rate at ${\big(}{\mathcal{A}_1^1},{\mathcal{A}_2^3}{\big)}$. As the total number of actions increases, the convergence time slots of DeepIOS increase because the agent needs to explore more possibilities to obtain a sufficiently effective policy. However, more optional actions do not necessarily yield higher data rate. On the one hand, when the phase-shift increment set $\mathcal{A}_1$ is fixed (i.e., $\mathcal{A}_1=\mathcal{A}_1^1$), the data rate of DeepIOS (with the action branch architecture) increases with the size of the reflecting amplitude set $\mathcal{A}_2$. Specifically, (i) from $\mathcal{A}_2^1$ to $\mathcal{A}_2^2$, the data rate gain of DeepIOS is 0.32\%, while the required convergence time slots remain unchanged, i.e., 0.00\%; and (ii) from $\mathcal{A}_2^2$ to $\mathcal{A}_2^3$, the data rate gain of DeepIOS is 1.03\%, while the required convergence time slots increase 1.33\%. Thus, among different types of reflecting amplitude sets, $\mathcal{A}_2^2$ is the most cost-effective. On the other hand, when the reflecting amplitude set $\mathcal{A}_2^1$ is fixed, the data rate of DeepIOS reaches the highest when the size of $\mathcal{A}_1$ is the smallest, i.e., $\mathcal{A}_1=\mathcal{A}_1^1$. Also, DeepIOS requires fewer time slots to converge in $\mathcal{A}_1^1$ than in $\mathcal{A}_1^2$ and $\mathcal{A}_1^3$. Hence, $\mathcal{A}_1^1$ is the best choice among different types of phase-shift increment sets.

The above results demonstrate the importance of the sub-action set design: a well-design sub-action set with an appropriate size is better than the over-small or over-large size. First, the size of the phase-shift increment set $\mathcal{A}_1$ can be small, but it should include: $\mathbf{w}(0)$ that keeps the phase-shift unchanged, $\mathbf{w}(-1/M)$ and $\mathbf{w}(1/M)$ that enable the agent to quickly correct from a negative action, and $\mathbf{w}(-3/M)$ and $\mathbf{w}(3/M)$ that accelerate the transition among shift-phases. Second, the larger the size of the reflecting amplitude set $\mathcal{A}_2$ is, the higher the data rate that DeepIOS can obtain. However, in terms of both data rate and convergence speed, a moderate reflecting amplitude set $\mathcal{A}_2$ is sufficient. Overall, $(A_1^1,A_2^2)$ is the most efficient phase-shift increment and reflecting amplitude combination for IOS configurations in MU-MIMO systems.


Fig. \ref{2a} also verifies the effectiveness of the action branch architecture. Specifically, when the size of the sub-action set is small, e.g., $\mathcal{A}_1={\mathcal{A}_1^1}$ and $\mathcal{A}_2={\mathcal{A}_2^1}$, the action branch architecture has no impact on DeepIOS's performance. As the size of the sub-action set gradually increases, the action branch architecture significantly improves the data rate and convergence speed of DeepIOS. When $\mathcal{A}_1={\mathcal{A}_1^3}$ and $\mathcal{A}_2={\mathcal{A}_2^1}$, compared with DeepIOS without the action branch architecture, the data rate and convergence speed gains of DeepIOS are 1.8\% and 9.3\%, respectively.

\textit{2) Case of MS-IOS:} the following five types of sub-action sets are considered, i.e., $({\mathcal{A}_1^1},{\mathcal{A}_2^1})$, $({\mathcal{A}_1^1},{\mathcal{A}_2^2})$, $({\mathcal{A}_1^1},{\mathcal{A}_2^3})$, $({\mathcal{A}_1^2},{\mathcal{A}_2^1})$, $({\mathcal{A}_1^3},{\mathcal{A}_2^1})$, where ${\mathcal{A}_1^1}$, ${\mathcal{A}_1^2}$, and ${\mathcal{A}_1^3}$ are the same as in Section \ref{Exp2}-1. The sizes of ${\mathcal{A}_2^1}$, ${\mathcal{A}_2^2}$, and ${\mathcal{A}_2^3}$ are 5, 10, and 15, respectively; and each of which sets ${\beta}_{\text{r},m}$, $\forall\,\,m{\in}\mathcal{M}$ to be 0 or 1 in random. As presented in Fig. \ref{2b} under MS-IOS, the performance of DeepIOS is also affected by the size of the sub-action set, and the action branch architecture enhances its performance. On the other hand, due to the inflexible control strategy between reflection and refraction, the data rate of DeepIOS under MS-IOS is lower than that under ES-IOS. To be specific, compared with the ES protocol, the MS protocol reduces the data rate of DeepIOS by about 23.7\%.

\begin{figure*}[t]
	\centering
    \subfloat[\label{3a} Sum data rate.]{
    \begin{minipage}[b]{0.326\textwidth}
	\includegraphics[scale=0.396]{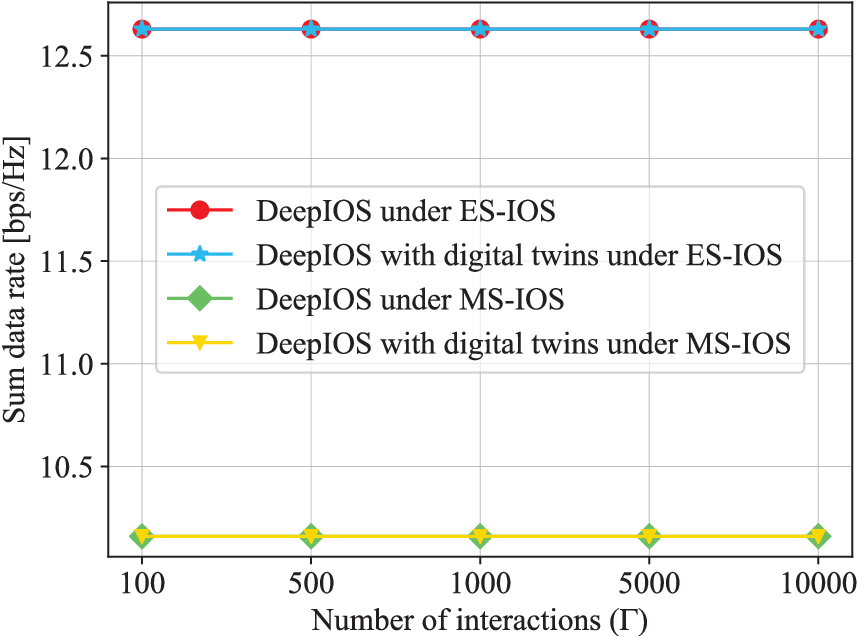}
    \end{minipage}}
    \subfloat[\label{3b} Time slots required for convergence.]{
    \begin{minipage}[b]{0.326\textwidth}
	\includegraphics[scale=0.396]{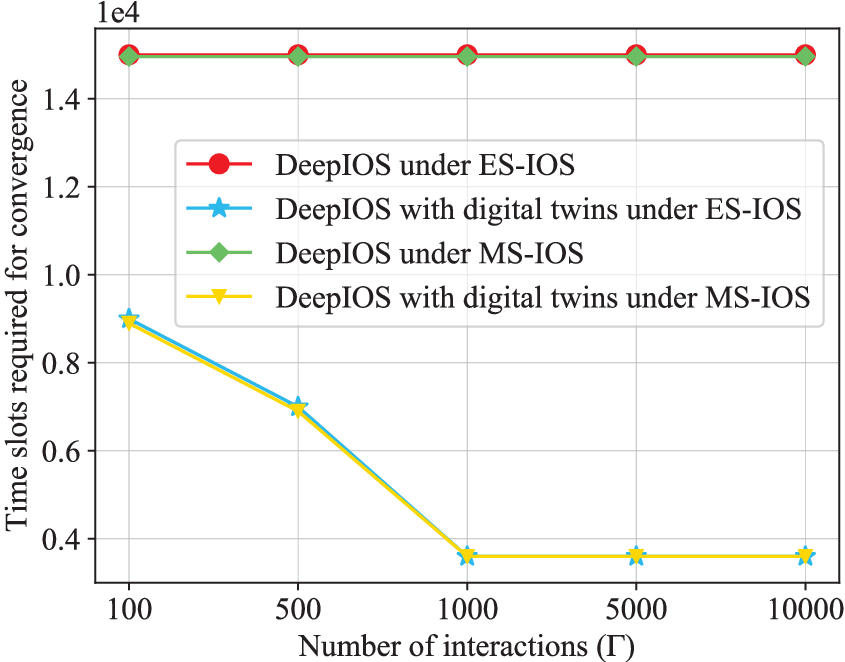}
    \end{minipage}}
    \subfloat[\label{3c} Simulation run-time.]{
    \begin{minipage}[b]{0.326\textwidth}
	\includegraphics[scale=0.396]{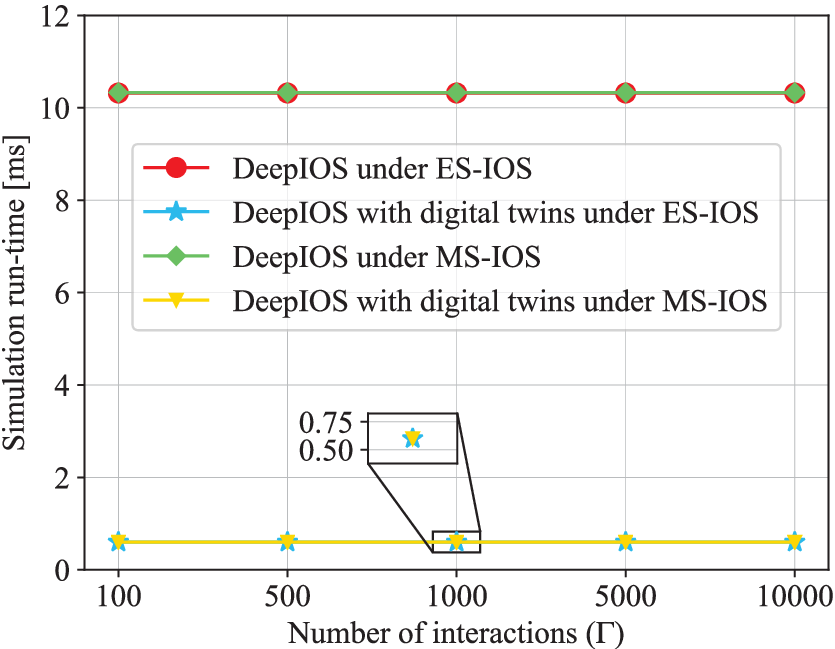}
    \end{minipage}}
	\caption{Sum data rates, time slots required for convergence, and simulation run-times of various schemes under both ES-IOS and MS-IOS cases.}
	\label{Fig3}
\end{figure*}

\subsection{Impact of Different Interaction Numbers in Digital Space}\label{Exp3}
In DeepIOS with digital twins, the digital DeepIOS interacts with the digital twin module to generate experience samples for DNN training, whilst the physical DeepIOS configures real-time IOS parameters in realistic systems. To evaluate the impact of different numbers of interactions in digital space on the physical DeepIOS's performance, this subsection considers five types of interaction numbers: $\Gamma=100$, $\Gamma=500$, $\Gamma=1000$, $\Gamma=5000$, and $\Gamma=10000$. The Rician factor $\lambda$ is 10. Furthermore, DeepIOS with digital twins only considers the physical DeepIOS's simulation run-time because only this module operates in the real environment.

\textit{1) Case of ES-IOS:} the phase-shift increment set ${\mathcal{A}}_1$ and the reflecting amplitude set ${\mathcal{A}}_2$ are the same as in Section \ref{Exp1}-1. Fig. \ref{3a}, Fig. \ref{3b}, and Fig. \ref{3c} provide various schemes' data rate, convergence time, and simulation run-time, respectively. As can be seen, when $\Gamma$ increases from 100 to 1000, the time slots required by the physical DeepIOS module to converge gradually decrease; however, when $\Gamma$ is further increased, the convergence time slots no longer decrease. This observation signifies that an appropriate $\Gamma$ is better than the over-small or over-large $\Gamma$. Although different $\Gamma$ will affect the convergence speed, the physical DeepIOS module can always converge to an effective strategy. On the other hand, owing to the digital twin module, the physical DeepIOS module only needs to perform IOS parameter configuration without training the DNN. Therefore, compared with DeepIOS without digital twins, DeepIOS with digital twins saves a large amount of simulation run-time---about 94.2\%. More importantly, the physical DeepIOS module determines an IOS parameter configuration in only 0.6ms, which guarantees the real-time demand in wireless communication systems.

\textit{2) Case of MS-IOS:} the phase-shift increment set ${\mathcal{A}}_1$ and the reflecting amplitude set ${\mathcal{A}}_2$ are the same as in Section \ref{Exp1}-2. Under MS-IOS, the convergence time slots and simulation run-time required by DeepIOS (with digital twins) are close to those under ES-IOS, as provided in Fig. \ref{3b} and Fig. \ref{3c}. As shown in Fig. \ref{3a}, the data rate of DeepIOS (with digital twins) under MS-IOS is 17.8\% lower than that under ES-IOS.

\subsection{Impact of Different Penalty Factors in the Reward Function}\label{Exp4}
Previous evaluations all assumed that the penalty factor $\omega$ in the reward function is 20. We now study the impact of different $\omega$ on the performance of DeepIOS, and illustrate the reason for setting $\omega=20$ in previous simulations. In particular, five types of penalty factors are considered: $\omega=5$, $\omega=10$, $\omega=15$, $\omega=20$, and $\omega=25$. The Rician factor $\lambda$ is 10. Under the ES-IOS case, the phase-shift increment set ${\mathcal{A}}_1$ and the reflecting amplitude set ${\mathcal{A}}_2$ are the same as in Section \ref{Exp1}-1; and under the MS-IOS case, the phase-shift increment set ${\mathcal{A}}_1$ and the reflecting amplitude set ${\mathcal{A}}_2$ are the same as in Section \ref{Exp1}-2.

TABLE \ref{tab:table4} summarizes the sum data rate of DeepIOS under different $\omega$ in both ES-IOS and MS-IOS cases. It can be found that as $\omega$ increases from 5 to 10, the sum data rate achieved by DeepIOS gradually increases: the data rate gain is more than 0.25bps/Hz. As $\omega$ further increases to 15 and 20, the performance of DeepIOS continues to be improved. However, when $\omega$ increases from 20 to 25, the performance of DeepIOS hardly changes. These observations indicate that $\omega=20$ is sufficient to enable the agent to learn an efficient IOS configuration policy. Therefore, we adopt $\omega=20$ in the reward function of DeepIOS throughout this paper.

\begin{table}[]
    \centering
    \caption{Sum data rate [bps/Hz] of DeepIOS under different $\omega$ in both ES-IOS and MS-IOS cases.}
    \begin{tabular}{|c|c|c|c|c|c|}\hline
     $\omega$ & 5 & 10 & 15 & 20 & 25\\\hline
     Sum data rate in ES-IOS & 12.20 & 12.45 & 12.56 & 12.63 & 12.63\\\hline
     Sum data rate in MS-IOS & 9.72 & 9.99 & 10.09 & 10.16 & 10.16\\\hline
    \end{tabular}
    \label{tab:table4}
\end{table}

\section{Conclusion}\label{5}
This paper put forth a new framework that integrates DRL and digital twins to enable model-free and real-time IOS configuration in MU-MIMO systems. First, by reformulating the sum-rate maximization problem into a POMDP, the DeepIOS scheme was developed to jointly optimize the phase-shift and amplitude of IOS, without prior sub-channels' CSI and UEs' mobility, Second, to reduce the computational complexity, DeepIOS introduced an action branch architecture, which optimizes different variables in a parallel and separate fashion. Besides, to guarantee the decision-making's real-time, this work built a digital twin module through supervised learning as a pre-verification platform for DeepIOS. The constructed framework is a closed-loop control system, where the digital DeepIOS interacts with the digital twin module to generate massive experiences for DeepIOS training; the physical DeepIOS controls IOS to improve the sum data rate; the digital DeepIOS delivers the DNN's parameter vector to the physical DeepIOS for policy improvement; and the data acquisition module collects real data from the practical MU-MIMO system to calibrate the digital twin module. Simulation results demonstrated that the proposed scheme (i) yields higher data rate than other schemes, (ii) satisfies the decision-making's real-time, (iii) maintains good adaptability to the environmental dynamics, and (iv) speeds up the rollout of IOS.

In this paper, the digital twin model is not considered in the formulated problem. Actually, the setup and implementation of digital twins will result in additional computational cost. It will be interesting to extend the current work to the joint communication and computation optimization framework. Besides, apart from the beamforming of IOS, the beamforming of BS also affects the sum data rate, especially in the presence of LoS. Therefore, in future work, we will further extend DeepIOS to jointly optimize the beamforming of BS and IOS.

\bibliography{reference}

\begin{thebibliography}{10}
\providecommand{\url}[1]{#1}
\csname url@samestyle\endcsname
\providecommand{\newblock}{\relax}
\providecommand{\bibinfo}[2]{#2}
\providecommand{\BIBentrySTDinterwordspacing}{\spaceskip=0pt\relax}
\providecommand{\BIBentryALTinterwordstretchfactor}{4}
\providecommand{\BIBentryALTinterwordspacing}{\spaceskip=\fontdimen2\font plus
\BIBentryALTinterwordstretchfactor\fontdimen3\font minus \fontdimen4\font\relax}
\providecommand{\BIBforeignlanguage}[2]{{%
\expandafter\ifx\csname l@#1\endcsname\relax
\typeout{** WARNING: IEEEtran.bst: No hyphenation pattern has been}%
\typeout{** loaded for the language `#1'. Using the pattern for}%
\typeout{** the default language instead.}%
\else
\language=\csname l@#1\endcsname
\fi
#2}}
\providecommand{\BIBdecl}{\relax}
\BIBdecl

\bibitem{wu2021intelligent}
Q.~Wu, S.~Zhang, B.~Zheng, C.~You, and R.~Zhang, ``Intelligent reflecting surface-aided wireless communications: A tutorial,'' \emph{IEEE Trans. Commun.}, vol.~69, no.~5, pp. 3313--3351, Jan. 2021.

\bibitem{huang2023rate}
R.~Huang, V.~W. Wong, and R.~Schober, ``Rate-splitting for intelligent reflecting surface-aided multiuser {VR} streaming,'' \emph{IEEE J. Sel. Areas Commun.}, vol.~41, no.~5, pp. 1516--1535, Jan. 2023.

\bibitem{wang2022intelligent}
W.~Wang and W.~Zhang, ``Intelligent reflecting surface configurations for smart radio using deep reinforcement learning,'' \emph{IEEE J. Sel. Areas Commun.}, vol.~40, no.~8, pp. 2335--2346, Jun. 2022.

\bibitem{xu2023intelligent}
S.~Xu, C.~Chen, Y.~Du, J.~Wang, and J.~Zhang, ``Intelligent reflecting surface backscatter enabled uplink coordinated multi-cell {MIMO} network,'' \emph{IEEE Trans. Wireless Commun.}, vol.~22, no.~8, pp. 5685--5696, Jan. 2023.

\bibitem{yang2021energy}
Z.~Yang, M.~Chen, W.~Saad, W.~Xu, M.~Shikh-Bahaei, H.~V. Poor, and S.~Cui, ``Energy-efficient wireless communications with distributed reconfigurable intelligent surfaces,'' \emph{IEEE Trans. Wireless Commun.}, vol.~21, no.~1, pp. 665--679, Jul. 2021.

\bibitem{niu2023active}
H.~Niu, Z.~Lin, K.~An, J.~Wang, G.~Zheng, N.~Al-Dhahir, and K.-K. Wong, ``Active {RIS} assisted rate-splitting multiple access network: Spectral and energy efficiency tradeoff,'' \emph{IEEE J. Sel. Areas Commun.}, vol.~41, no.~5, pp. 1452--1467, Jun. 2023.

\bibitem{aung2023energy}
P.~S. Aung, Y.~M. Park, Y.~K. Tun, Z.~Han, and C.~S. Hong, ``Energy-efficient communication networks via multiple aerial reconfigurable intelligent surfaces: {DRL} and optimization approach,'' \emph{IEEE Trans. Veh. Tech.}, vol.~73, no.~3, pp. 4277--4292, Oct. 2023.

\bibitem{yuan2020intelligent}
J.~Yuan, Y.-C. Liang, J.~Joung, G.~Feng, and E.~G. Larsson, ``Intelligent reflecting surface-assisted cognitive radio system,'' \emph{IEEE Trans. Commun.}, vol.~69, no.~1, pp. 675--687, Oct. 2020.

\bibitem{pan2020intelligent}
C.~Pan, H.~Ren, K.~Wang, M.~Elkashlan, A.~Nallanathan, J.~Wang, and L.~Hanzo, ``Intelligent reflecting surface aided {MIMO} broadcasting for simultaneous wireless information and power transfer,'' \emph{IEEE J. Sel. Areas Commun.}, vol.~38, no.~8, pp. 1719--1734, Jun. 2020.

\bibitem{zeng2021reconfigurable}
S.~Zeng, H.~Zhang, B.~Di, Y.~Tan, Z.~Han, H.~V. Poor, and L.~Song, ``Reconfigurable intelligent surfaces in {6G}: Reflective, transmissive, or both?'' \emph{IEEE Commun. Lett.}, vol.~25, no.~6, pp. 2063--2067, Feb. 2021.

\bibitem{zhang2022intelligent1}
H.~Zhang and B.~Di, ``Intelligent omni-surfaces: Simultaneous refraction and reflection for full-dimensional wireless communications,'' \emph{IEEE Commun. Surveys Tuts.}, vol.~24, no.~4, pp. 1997--2028, Aug. 2022.

\bibitem{fang2023intelligent}
S.~Fang, G.~Chen, P.~Xiao, K.-K. Wong, and R.~Tafazolli, ``Intelligent omni surface-assisted self-interference cancellation for full-duplex {MISO} system,'' \emph{IEEE Trans. Wireless Commun.}, vol.~23, no.~3, pp. 2268--2281, Jul. 2023.

\bibitem{zhang2022intelligent2}
H.~Zhang, S.~Zeng, B.~Di, Y.~Tan, M.~Di~Renzo, M.~Debbah, Z.~Han, H.~V. Poor, and L.~Song, ``Intelligent omni-surfaces for full-dimensional wireless communications: Principles, technology, and implementation,'' \emph{IEEE Commun. Mag.}, vol.~60, no.~2, pp. 39--45, Feb. 2022.

\bibitem{zhang2022meta}
Y.~Zhang, B.~Di, H.~Zhang, Z.~Han, H.~V. Poor, and L.~Song, ``Meta-wall: Intelligent omni-surfaces aided multi-cell {MIMO} communications,'' \emph{IEEE Trans. Wireless Commun.}, vol.~21, no.~9, pp. 7026--7039, Mar. 2022.

\bibitem{liu2022full}
Y.~Liu, B.~Duo, Q.~Wu, X.~Yuan, and Y.~Li, ``Full-dimensional rate enhancement for {UAV}-enabled communications via intelligent omni-surface,'' \emph{IEEE Wireless Commun. Lett.}, vol.~11, no.~9, pp. 1955--1959, Jul. 2022.

\bibitem{zhang2021intelligent}
S.~Zhang, H.~Zhang, B.~Di, Y.~Tan, M.~Di~Renzo, Z.~Han, H.~V. Poor, and L.~Song, ``Intelligent omni-surfaces: Ubiquitous wireless transmission by reflective-refractive metasurfaces,'' \emph{IEEE Trans. Wireless Commun.}, vol.~21, no.~1, pp. 219--233, Jul. 2021.

\bibitem{chen2022robust}
Y.~Chen, Y.~Wang, Z.~Wang, and P.~Zhang, ``Robust beamforming for active reconfigurable intelligent omni-surface in vehicular communications,'' \emph{IEEE J. Sel. Areas Commun.}, vol.~40, no.~10, pp. 3086--3103, Aug. 2022.

\bibitem{cai2022joint}
W.~Cai, M.~Li, Y.~Liu, Q.~Wu, and Q.~Liu, ``Joint beamforming design for intelligent omni surface assisted wireless communication systems,'' \emph{IEEE Trans. Wireless Commun.}, vol.~22, no.~2, pp. 1281--1297, Sept. 2022.

\bibitem{wang2022intelligentomni}
W.~Wang, W.~Ni, H.~Tian, and L.~Song, ``Intelligent omni-surface enhanced aerial secure offloading,'' \emph{IEEE Trans. Veh. Tech.}, vol.~71, no.~5, pp. 5007--5022, Feb. 2022.

\bibitem{fang2022intelligent}
S.~Fang, G.~Chen, Z.~Abdullah, and Y.~Li, ``Intelligent omni surface-assisted secure {MIMO} communication networks with artificial noise,'' \emph{IEEE Commun. Lett.}, vol.~26, no.~6, pp. 1231--1235, Mar. 2022.

\bibitem{benaya2023physical}
A.~Benaya, M.~H. Ismail, A.~S. Ibrahim, and A.~A. Salem, ``Physical layer security enhancement via intelligent omni-surfaces and {UAV}-friendly jamming,'' \emph{IEEE Access}, vol.~11, pp. 2531--2544, Jan. 2023.

\bibitem{wang2022safeguarding}
W.~Wang, W.~Ni, H.~Tian, Z.~Yang, C.~Huang, and K.-K. Wong, ``Safeguarding {NOMA} networks via reconfigurable dual-functional surface under imperfect {CSI},'' \emph{IEEE J. Sel. Topics Signal Process.}, vol.~16, no.~5, pp. 950--966, May 2022.

\bibitem{zhang2022dual}
Y.~Zhang, B.~Di, H.~Zhang, M.~Dong, L.~Yang, and L.~Song, ``Dual codebook design for intelligent omni-surface aided communications,'' \emph{IEEE Trans. Wireless Commun.}, vol.~21, no.~11, pp. 9232--9245, May 2022.

\bibitem{mnih2015human}
V.~Mnih, K.~Kavukcuoglu, D.~Silver, A.~A. Rusu, J.~Veness, M.~G. Bellemare, A.~Graves, M.~Riedmiller, A.~K. Fidjeland, G.~Ostrovski \emph{et~al.}, ``Human-level control through deep reinforcement learning,'' \emph{nature}, vol. 518, no. 7540, pp. 529--533, Feb. 2015.

\bibitem{barricelli2019survey}
B.~R. Barricelli, E.~Casiraghi, and D.~Fogli, ``A survey on digital twin: Definitions, characteristics, applications, and design implications,'' \emph{IEEE access}, vol.~7, pp. 167\,653--167\,671, Nov. 2019.

\bibitem{sutton2018reinforcement}
R.~S. Sutton and A.~G. Barto, \emph{Reinforcement learning: An introduction}.\hskip 1em plus 0.5em minus 0.4em\relax Cambridge, MA, USA:MIT press, 2018.

\bibitem{nguyen2021digital}
H.~X. Nguyen, R.~Trestian, D.~To, and M.~Tatipamula, ``Digital twin for {5G} and beyond,'' \emph{IEEE Commun. Mag.}, vol.~59, no.~2, pp. 10--15, Feb. 2021.

\bibitem{wu2021digital}
Y.~Wu, K.~Zhang, and Y.~Zhang, ``Digital twin networks: A survey,'' \emph{IEEE Internet Things J.}, vol.~8, no.~18, pp. 13\,789--13\,804, May 2021.

\bibitem{zhao2022simultaneously}
J.~Zhao, Y.~Zhu, X.~Mu, K.~Cai, Y.~Liu, and L.~Hanzo, ``Simultaneously transmitting and reflecting reconfigurable intelligent surface ({STAR-RIS}) assisted {UAV} communications,'' \emph{IEEE J. Sel. Areas Commun.}, vol.~40, no.~10, pp. 3041--3056, Aug. 2022.

\bibitem{adhikary2023artificial}
A.~Adhikary, M.~S. Munir, A.~D. Raha, Y.~Qiao, S.~H. Hong, E.-N. Huh, and C.~S. Hong, ``An artificial intelligence framework for holographic beamforming: Coexistence of holographic {MIMO} and intelligent omni-surface,'' in \emph{IEEE ICOIN}, Jan. 2023, pp. 19--24.

\bibitem{luo2024meta}
Q.~Luo, Z.~Han, and B.~Di, ``Meta-critic reinforcement learning for intelligent omnidirectional surface assisted multi-user communications,'' \emph{IEEE Trans. Wireless Commun.}, vol.~23, no.~8, pp. 9085--9098, Feb. 2024.

\bibitem{fan2021digital}
B.~Fan, Y.~Wu, Z.~He, Y.~Chen, T.~Q. Quek, and C.-Z. Xu, ``Digital twin empowered mobile edge computing for intelligent vehicular lane-changing,'' \emph{IEEE Netw.}, vol.~35, no.~6, pp. 194--201, Nov./Dec. 2021.

\bibitem{huang2024digital}
X.~Huang, H.~Yang, S.~Hu, and X.~Shen, ``Digital twin-driven network architecture for video streaming,'' \emph{IEEE Netw.}, Apr. 2024, doi: 10.1109/MNET.2024.3386030.

\bibitem{peng2024stochastic}
Y.~Peng, J.~Duan, J.~Zhang, W.~Li, Y.~Liu, and F.~Jiang, ``Stochastic long-term energy optimization in digital twin-assisted heterogeneous edge networks,'' \emph{IEEE J. Sel. Areas Commun.}, Jul. 2024, doi: 10.1109/JSAC.2024.3431581.

\bibitem{DeepLearning}
I.~Goodfellow, Y.~Bengio, and A.~Courville, \emph{Deep learning}.\hskip 1em plus 0.5em minus 0.4em\relax Cambridge, MA, USA:MIT press, 2016.

\bibitem{tavakoli2018action}
A.~Tavakoli, F.~Pardo, and P.~Kormushev, ``Action branching architectures for deep reinforcement learning,'' in \emph{Proc. AAAI}, vol.~32, no.~1, 2018.

\bibitem{caruana2006empirical}
R.~Caruana and A.~Niculescu-Mizil, ``An empirical comparison of supervised learning algorithms,'' in \emph{ACM ICML}, 2006, pp. 161--168.

\bibitem{qian2020noma}
L.~Qian, Y.~Wu, F.~Jiang, N.~Yu, W.~Lu, and B.~Lin, ``{NOMA} assisted multi-task multi-access mobile edge computing via deep reinforcement learning for industrial internet of things,'' \emph{IEEE Trans. Ind. Informat.}, vol.~17, no.~8, pp. 5688--5698, Aug. 2020.

\bibitem{cho2014learning}
K.~Cho, B.~Van~Merri{\"e}nboer, C.~Gulcehre, D.~Bahdanau, F.~Bougares, H.~Schwenk, and Y.~Bengio, ``Learning phrase representations using {RNN} encoder-decoder for statistical machine translation,'' in \emph{EMNLP}, Sept. 2014, pp. 1724--1734.

\bibitem{ye2021multi}
{X. Ye, Y. Yu, and L. Fu}, ``Multi-channel opportunistic access for heterogeneous networks based on deep reinforcement learning,'' \emph{IEEE Trans. Wireless Commun.}, vol.~21, no.~2, pp. 794--807, Feb. 2022.

\bibitem{mihai2022digital}
S.~Mihai, M.~Yaqoob, D.~V. Hung, W.~Davis, P.~Towakel, M.~Raza, M.~Karamanoglu, B.~Barn, D.~Shetve, R.~V. Prasad \emph{et~al.}, ``Digital twins: A survey on enabling technologies, challenges, trends and future prospects,'' \emph{IEEE Commun. Surveys Tuts.}, vol.~24, no.~4, pp. 2255--2291, Sept. 2022.

\bibitem{he2016deep}
K.~He, X.~Zhang, S.~Ren, and J.~Sun, ``Deep residual learning for image recognition,'' in \emph{IEEE CVPR}, Jun. 2016, pp. 770--778.

\bibitem{ye2020deep}
X.~Ye, Y.~Yu, and L.~Fu, ``Deep reinforcement learning based {MAC} protocol for underwater acoustic networks,'' \emph{IEEE Trans. Mobile Comput.}, vol.~21, no.~5, pp. 1625--1638, May 2022.

\bibitem{keras}
\BIBentryALTinterwordspacing
F.~Chollet, \emph{Keras: {The} python deep learning library}, Jun. 2018. [Online]. Available: \url{https://keras.io.}
\BIBentrySTDinterwordspacing

\bibitem{pyitpp}
\BIBentryALTinterwordspacing
V.~Saxena, \emph{py-itpp}, 2020. [Online]. Available: \url{https://github.com/vidits-kth/py-itpp.}
\BIBentrySTDinterwordspacing

\bibitem{tabassum2019fundamentals}
H.~Tabassum, M.~Salehi, and E.~Hossain, ``Fundamentals of mobility-aware performance characterization of cellular networks: {A} tutorial,'' \emph{IEEE Commun. Surveys Tuts.}, vol.~21, no.~3, pp. 2288--2308, Mar. 2019.

\end{thebibliography}

\bibliographystyle{IEEEtran}
\begin{IEEEbiography}[{\includegraphics[width=1in,height=1.25in,clip,keepaspectratio]{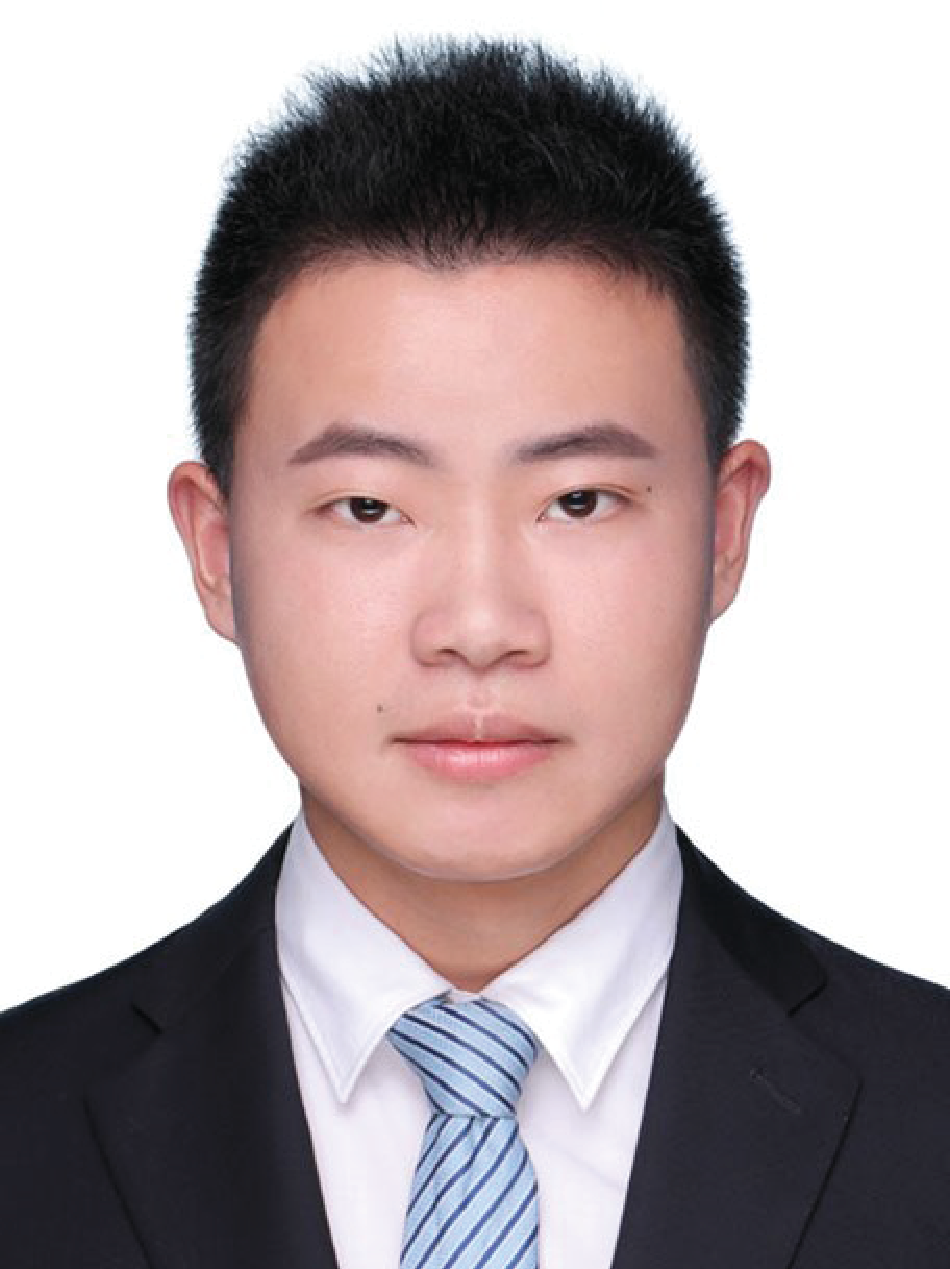}}]{Xiaowen Ye} is a post-doctoral research fellow with the Department of Electrical Engineering, City University of Hong Kong, Hong Kong, China. He received the Ph.D. degree in communication and information systems from Xiamen University, Xiamen, China, in 2024. His research interests include deep reinforcement learning, wireless network optimization, and dynamic resource allocation.
\end{IEEEbiography}

\begin{IEEEbiography}[{\includegraphics[width=1in,height=1.25in]{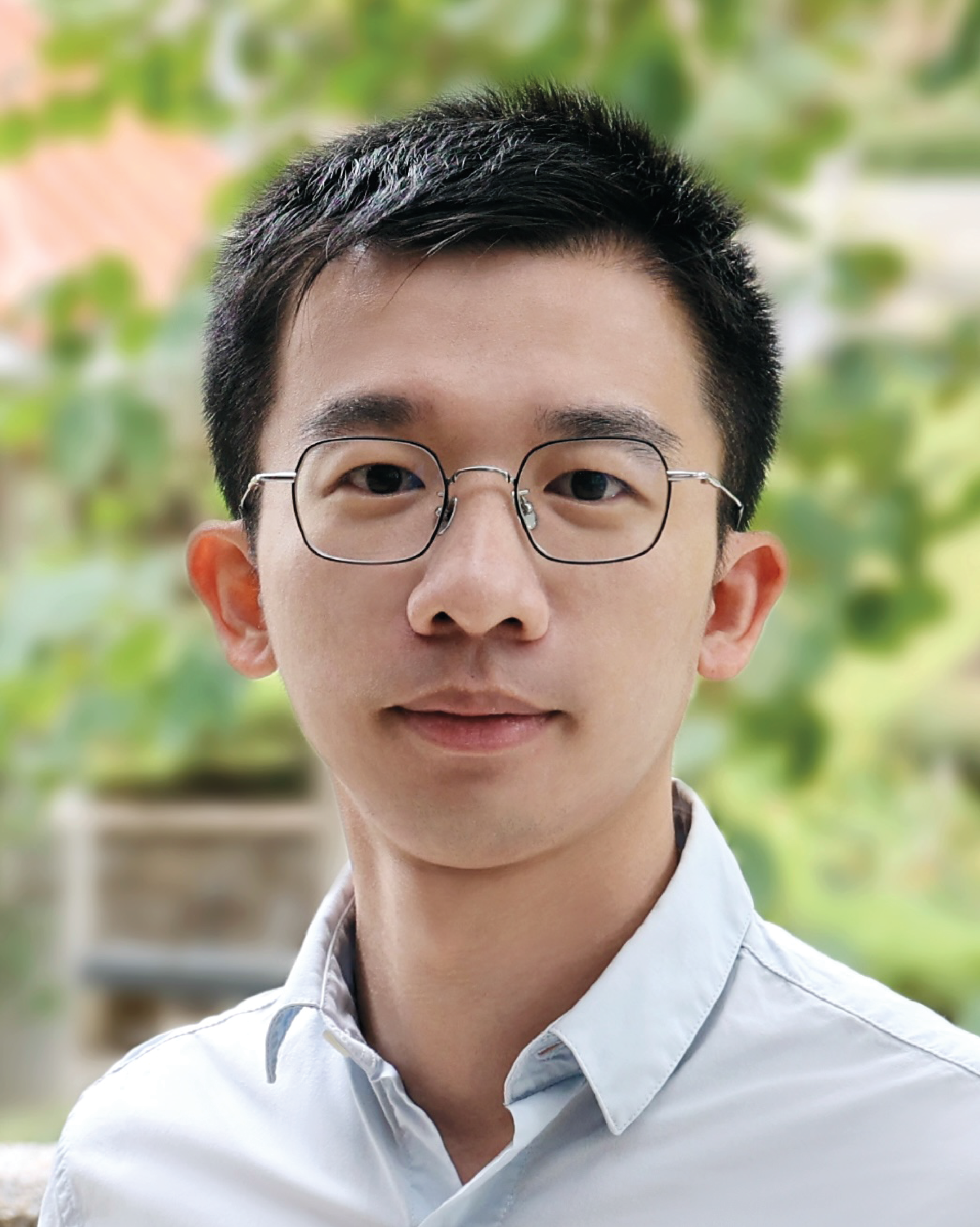}}]{Xianghao Yu} (Senior Member, IEEE) received his B.Eng. degree in information engineering from Southeast University, Nanjing, China, in 2014, and his Ph.D. degree in electronic and computer engineering from the Hong Kong University of Science and Technology (HKUST), Hong Kong, China, in 2018.
He is currently an Assistant Professor with the Department of Electrical Engineering at City University of Hong Kong (CityU), Hong Kong, China. From 2018 - 2020, he was a Humboldt Post-Doctoral Research Fellow with the Institute for Digital Communications, Friedrich-Alexander University of Erlangen-Nuremberg (FAU), Erlangen, Germany. Before joining CityU, he was a Research Assistant Professor with the Department of Electronic and Computer Engineering at HKUST. His research interests include intelligent reflecting surface-assisted communications, integrated sensing and communications, near-field communications, and wireless artificial intelligence.
Dr. Yu has co-authored the book \emph{Stochastic Geometry Analysis of Multi-Antenna Wireless Networks} (Springer, 2019). He was listed as the World's Top 2\% Scientist by Stanford University from 2020 to 2022. Dr. Yu received the IEEE Global Communications Conference (GLOBECOM) 2017 Best Paper Award, the 2018 IEEE Signal Processing Society Young Author Best Paper Award, the IEEE GLOBECOM 2019 Best Paper Award, the 2023 IEEE Communications Society Leonard G. Abraham Prize, and the 2024 IEEE ComSoc Asia-Pacific Outstanding Young Researcher Award. He was also recognized as an Exemplary Reviewer of the \textsc{IEEE Transactions on Wireless Communications} in 2017 and 2018, and an Exemplary Reviewer of the \textsc{IEEE Transactions on Communications} in 2021 and 2022. He is an Editor of \textsc{IEEE Transactions on Mobile Computing} and \textsc{IEEE Communications Letters}.
\end{IEEEbiography}

\begin{IEEEbiography}[{\includegraphics[width=1in,height=1.25in,clip,keepaspectratio]{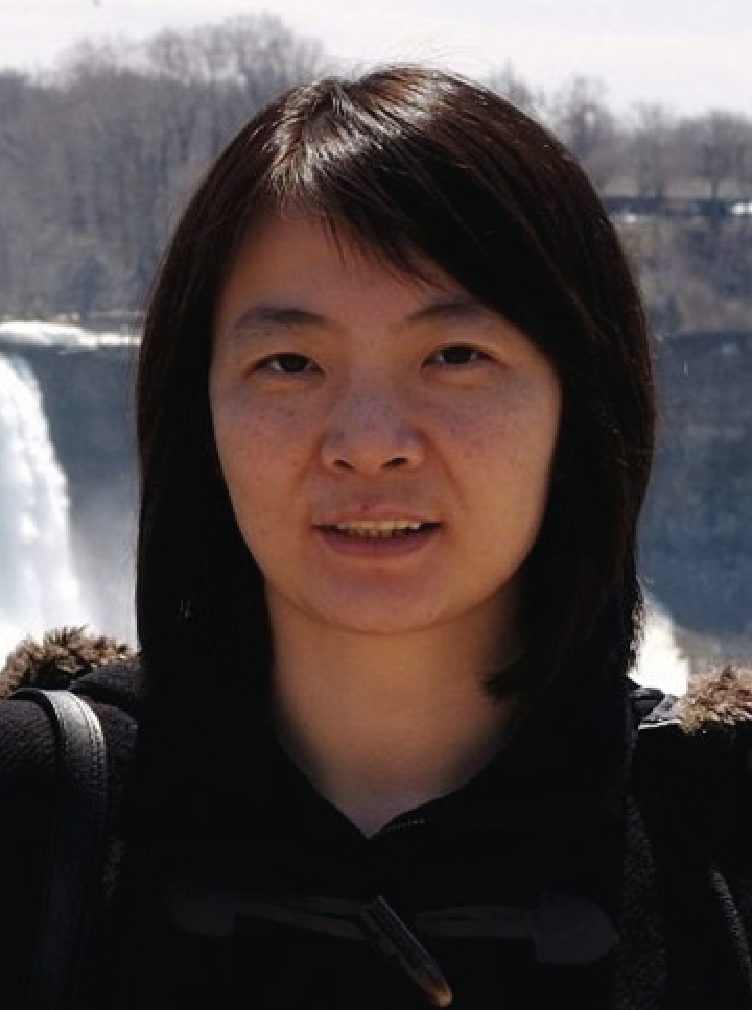}}]{Liqun Fu} (S'08-M'11-SM'17) is a Full Professor of the School of Informatics at Xiamen University, China. She received her Ph.D. Degree in Information Engineering from The Chinese University of Hong Kong in 2010. She was a post-doctoral research fellow with the Institute of Network Coding of The Chinese University of Hong Kong, and the ACCESS Linnaeus Centre of KTH Royal Institute of Technology during 2011-2013 and 2013-2015, respectively. She was with ShanghaiTech University as an Assistant Professor during 2015-2016. Her research interests are mainly in communication theory, optimization theory, game theory, and learning theory, with applications in wireless networks. She is on the editorial board of IEEE Communications Letters and the Journal of Communications and Information Networks (JCIN). She served as the Technical Program Co-Chair of IEEE/CIC ICCC 2021 and the GCCCN Workshop of the IEEE INFOCOM 2014, the Publicity Co-Chair of the GSNC Workshop of the IEEE INFOCOM 2016, and the Web Chair of the IEEE WiOpt 2018. She also serves as a TPC member for many leading conferences in communications and networking, such as the IEEE INFOCOM, ICC, and GLOBECOM.
\end{IEEEbiography}

\end{document}